\documentclass[iop]{emulateapj}
\newcommand{\etal}{et~al.}
\newcommand{\Msol}{M$_{\odot}$}

\newcommand{\cgsflux}{erg~s$^{-1}$~cm$^{-2}$}
\newcommand{\cgsflam}{erg~s$^{-1}$~cm$^{-2}$~\AA$^{-1}$}

\newcommand{\kms}{\hbox{km~s$^{-1}$}}
\newcommand{\cc}{\hbox{cm$^{-3}$}}

\newcommand{\ArII}{\hbox{{\rm Ar}\kern 0.1em{\sc ii}}}
\newcommand{\ArIII}{\hbox{{\rm Ar}\kern 0.1em{\sc iii}}}
\newcommand{\CIV}{\hbox{{\rm C}\kern 0.1em{\sc iv}}}
\newcommand{\HI}{\hbox{{\rm H}\kern 0.1em{\sc i}}}
\newcommand{\HII}{\hbox{{\rm H}\kern 0.1em{\sc ii}}}
\newcommand{\HeI}{\hbox{{\rm He}\kern 0.1em{\sc i}}}
\newcommand{\HeII}{\hbox{{\rm He}\kern 0.1em{\sc ii}}}
\newcommand{\NII}{\hbox{{\rm N}\kern 0.1em{\sc ii}}}
\newcommand{\OI}{\hbox{{\rm O}\kern 0.1em{\sc i}}}
\newcommand{\OII}{\hbox{{\rm O}\kern 0.1em{\sc ii}}}
\newcommand{\OIII}{\hbox{{\rm O}\kern 0.1em{\sc iii}}}
\newcommand{\OIIlong}{{\rm O}\kern 0.1em{\sc ii}~$\lambda 3727$} 
\newcommand{\FeII}{\hbox{{\rm Fe}\kern 0.1em{\sc ii}}}
\newcommand{\NeII}{\hbox{{\rm Ne}\kern 0.1em{\sc ii}}}
\newcommand{\NeIII}{\hbox{{\rm Ne}\kern 0.1em{\sc iii}}}
\newcommand{\NeV}{\hbox{{\rm Ne}\kern 0.1em{\sc v}}}
\newcommand{\SII}{\hbox{{\rm S}\kern 0.1em{\sc ii}}}
\newcommand{\SIII}{\hbox{{\rm S}\kern 0.1em{\sc iii}}}
\newcommand{\SIV}{\hbox{{\rm S}\kern 0.1em{\sc iv}}}
\newcommand{\SiIV}{\hbox{{\rm Si}\kern 0.1em{\sc iv}}}
\newcommand{\MgII}{\hbox{{\rm Mg}\kern 0.1em{\sc ii}}}
\newcommand{\Halpha}{\hbox{{\rm H}\kern 0.1em$\alpha$}}
\newcommand{\Hbeta}{\hbox{{\rm H}\kern 0.1em$\beta$}}
\newcommand{\Heopta}{\hbox{{\rm He}\kern 0.1em{\sc i}}~$6678$}
\newcommand{\Heoptb}{\hbox{{\rm He}\kern 0.1em{\sc i}}~$5876$}
\newcommand{\Heoptc}{\hbox{{\rm He}\kern 0.1em{\sc i}}~$4471$}
\newcommand{\Brgam}{\hbox{{\rm Br}\kern 0.1em$\gamma$}}
\newcommand{\Brten}{\hbox{{\rm Br}\kern 0.1em$10$}}
\newcommand{\Breleven}{\hbox{{\rm Br}\kern 0.1em$11$}}
\newcommand{\HeIh}{\hbox{{\rm He}\kern 0.1em{\sc i}}~$1.7$~{\micron}}
\newcommand{\HeIk}{\hbox{{\rm He}\kern 0.1em{\sc i}}~$2.06$~{\micron}}
\newcommand{\squishlist}{
   \begin{list}{$\bullet$}
    { \setlength{\itemsep}{0pt}      \setlength{\parsep}{1pt}
      \setlength{\topsep}{3pt}       \setlength{\partopsep}{0pt}
      \setlength{\leftmargin}{1.5em} \setlength{\labelwidth}{1em}
      \setlength{\labelsep}{0.5em} } }
\newcommand{\squishend}{
    \end{list}  }

\newcommand{\arcnamelong}{RCSGA 032727-132609}
\newcommand{\arcname}{RCS0327}

\slugcomment{ApJ in press}
\shorttitle{Physical conditions in a lensed z$=$1.7 galaxy}
\shortauthors{Rigby \etal}

\begin{document}
\title{The physical conditions of a lensed star-forming galaxy at z$=$1.7}
\author{J.~R.~Rigby\altaffilmark{1,2}, E. Wuyts\altaffilmark{3,4}, M.~D.~Gladders\altaffilmark{3,4}, 
K.~Sharon\altaffilmark{4}, and G.~D.~Becker\altaffilmark{5}}
\altaffiltext{1}{NASA Goddard Space Flight Center, Code 665, Greenbelt MD 20771}
\altaffiltext{2}{Previous:  Carnegie Fellow, Carnegie Institution for Science, 813 Santa Barbara St., Pasadena, CA 91101}
\altaffiltext{3}{Department of Astronomy \& Astrophysics, The University of Chicago, 5640 S.~Ellis Avenue, Chicago, IL 60637}
\altaffiltext{4}{Kavli Institute for Cosmological Physics, University of Chicago, 5640 S.~Ellis Avenue, Chicago, IL 60637}
\altaffiltext{5}{KICC Fellow, Kavli Institute for Cosmology, Cambridge England}

\email{Jane.R.Rigby@nasa.gov}

\begin{abstract}
We report rest-frame optical Keck/NIRSPEC spectroscopy of the bright lensed galaxy 
\arcnamelong\ at z=1.7037.  From precise measurements of the nebular lines, 
we infer a number of physical properties: 
redshift, extinction, star formation rate, ionization parameter, electron density, 
electron temperature, oxygen abundance, and N/O, Ne/O, and Ar/O abundance ratios.  
The limit on [O III]~4363~\AA\ tightly constrains the oxygen abundance via the ``direct'' or 
$T_e$ method, for the first time in an average--metallicity galaxy at z$\sim$2.  
We compare this result to several standard  ``bright-line'' O abundance diagnostics, thereby 
testing these empirically--calibrated diagnostics \textit{in situ}.  Finally, we explore the 
positions of lensed and unlensed galaxies in standard diagnostic diagrams, and explore the 
diversity of ionization conditions and mass--metallicity ratios at z$=$2.
\end{abstract}

\keywords{galaxies: high-redshift---galaxies: evolution---gravitational lensing}

\section{Introduction} 		\label{sec:intro}

Our knowledge of the Universe's star formation history
has advanced remarkably in the past dozen years.  
First rest-UV photometry \citep{madau96,madau98}, then 
24~\micron\ Spitzer photometry for  $0<z<2$ galaxies \citep{caputi07, lefloch05}
and rest-UV photometry at higher redshift (c.f. \citealt{ouchi09,bouwens10}), 
has shown us that the SFR rises steeply from z$=$0 to z$=$1, has a
broad plateau from z$=$1 to z$=$4--5, and falls off (with debated slope) out to reionization.  
Thus, the question, ``What is the star formation history of
the Universe?'' has been answered reasonably, and the focus
has shifted toward,  ``Which galaxies formed their stars when, and why, and 
how did that star-formation change those galaxies and their environments?''

While we have identified the galaxies that formed most of the Universe's stars, we
have much to learn about how that process occurred.
One key question is, what were the physical conditions inside these galaxies--metallicity, 
abundance, extinction, stellar effective temperature, 
electron temperature and density---compared to star-forming galaxies today?  
How did these physical conditions evolve through episodes of star formation and gas accretion?

Until the era of extremely large telescopes arrives, our best chance to address this question 
is to study galaxies that are highly magnified by gravitational lensing.  In such rare cases, 
magnification factors of 20--30 make diagnostic spectroscopy possible with current telescopes.  
Such work was pioneered in the galaxy MS1512-cB58 (\citealt{yee}; hereafter cB58)
 by \citet{pettini00} and \citet{teplitz}, and 
can now be extended to a larger sample thanks to  discoveries of bright lensed galaxies 
(e.g., \citealt{allam07,horseshoe,smail-cosmic-eye,rigby08,koester}.)

Among these new discoveries is  \arcnamelong, hereafter RCS0327, at z$=$1.7 \citep{wuyts}, 
discovered from the second Red Sequence Cluster Survey (RCS2; Gilbank \etal\ in prep.).
With an integrated $g$-band magnitude of 19.15, we believe \arcname\ to be the brightest 
high-redshift lensed galaxy yet found.
\citet{wuyts} present the discovery, spectroscopic confirmation, deep-follow-up imaging, 
spectral energy distribution modeling, and preliminary lensing analysis.
In this paper, using 1.3~hr of integration with NIRSPEC on Keck, we determine 
with unprecedented precision the physical conditions of star formation 
in this hopefully--typical galaxy at the crucial epoch of z$\sim$2.

\section{Methods}      		\label{sec:methods}
\begin{figure}[h]
\figurenum{1}
\includegraphics[width=3.4in,angle=0]{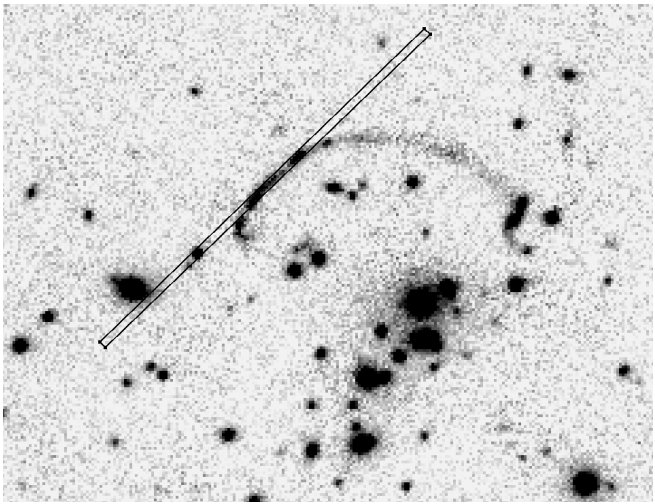}
\figcaption{Finderchart for \arcname.  This J-band image from \citet{wuyts} illustrates
how we positioned the NIRSPEC slit, which is  42 $\times$ 0.76\arcsec.  N is up and E is left.  
For simplicity we show the slit centered on the source; in fact it the source 
was placed on the left half and then the right half of the slit (``an AB pattern''), 
with the nods separated by 10--15\arcsec.}\label{fig:finderchart}
\end{figure}

\arcname\ was observed on 04 Feb.\ 2010 UT with the NIRSPEC spectrograph 
(McLean et al. 1998) on the Keck II telescope.  
The weather was clear, and the seeing was measured 
as 0.85\arcsec\ and 0.45\arcsec\ when the telescope was focused during the night.  
We used the low--resolution mode and the 0.76\arcsec\ $\times$ 42\arcsec\  longslit.  
We targeted the brightest $\sim10$\arcsec\ of the arc, at a position angle of 134\arcdeg, as shown 
in Figure~\ref{fig:finderchart}.
The target was acquired by offsetting from nearby stars on the near-IR slit-viewing camera, 
target acquisition was verified by direct imaging on this camera.
The target was nodded along the slit in an AB pattern,  with exposures of 600~s per nod.
Table~1 
summarizes the filters and exposure times.
The A0V star HD 23683 was observed every hour as a telluric standard.

We reduced the spectra using the \textit{nirspec\_reduce} package written by one of us 
(G.~D.~Becker), which 
uses lamp exposures to flatten the data, the sky lines to wavelength calibrate, 
and optimally fits and subtracts the sky following \citet{kelson}.  
For each frame, we measured the spatial profile of the lensed arc by fitting the
brightest emission lines, then used this spatial profile to optimally extract the spectrum.

The arc is extended over 38\arcsec, roughly 10\arcsec\ of which was captured by the NIRSPEC slit.  
In a subsequent paper, we will analyze the spatial variation of physical conditions across the arc.
In this paper, we consider the integrated spectrum.

\begin{table}
\label{tab:obslog}
\begin{center}
\caption{Observation Log}
\begin{tabular}{lll}
filter &   t(s)  & wavelength range(\micron)\\ 
\tableline\tableline
NIRSPEC-1     &  1200  &  0.948--1.16\\
NIRSPEC-3     &  2400  &  1.14--1.43\\
NIRSPEC-6     &  1200  &  1.76--2.19\\
\tableline
\end{tabular}
\end{center}
\end{table}

Each extracted spectrum was corrected for telluric absorption and fluxed using the 
tool \textit{xtellcor\_general} \citep{vacca}, using the closest--in--time 
observation of the AOV standard star.  The flux level is thus appropriate for the 
fraction of the galaxy inside the slit, not for the whole galaxy.
In \S\ref{sec:fraction} we estimate the factor by which our NIRSPEC fluxes should be 
scaled to represent the whole galaxy.

Using a telluric standard to flux the spectra provides excellent relative 
fluxing within a given filter, but because observations were taken in three separate filters, 
there can be offsets between filters due to differential slit losses, due for example to 
changes in seeing or pointing.   In addition, the lines in filter NIRSPEC-6 are observed right 
at the edge of an atmospheric transmission window, and thus may suffer especially high 
telluric variability.  We address these issues in \S\ref{sec:tweak}.

For each filter, we combined the individual flux--calibrated 1D spectra with a weighted average, 
producing one fluxed spectrum for each of the three filters.  

We fit line fluxes as follows.  
We fit each isolated line with a Gaussian to measure the line flux, using Levenberg--Marquardt fitting.
The continuum used was the mean flux in an adjacent spectral region, chosen by hand for each line.  
For partially--overlapping lines, we simultaneously fit multiple Gaussians using A.~Marble's 
implementation
of the IDL Levenberg-Marquardt least-squares fitting code MPFITFUN \citep{mpfitfun}.
We report line fluxes in Table~\ref{tab:fluxes}.

\begin{deluxetable}{llll}
\tablecolumns{5}
\tablewidth{0pc}
\tablecaption{Measured line fluxes for \arcname\label{tab:fluxes}}
\tablehead{
\colhead{line} &  \colhead{$\lambda_{obs}$} &  \colhead{flux} & \colhead{dflux}}
\startdata
\bf{Filter N1}\\
$[$O II$]$ 3727     & 1.00808       & 72.0  & 0.2\\
$[$Ne III$]$ 3869   & 1.046273      & 7.8   & 1.3\\
H8+HeI 3890         & 1.05195       & 6.6   & 0.4\\
H$\epsilon$         & 1.0729        & $<$3.8 & limit \\
H$\delta$           & 1.1093        & 7.1   & 1.7\\
\bf{Filter N3}\\
H$\gamma$           & 1.17385       & 13.5  & 2\\
$[$O III$]$ 4363    & 1.1800        & $<$1.5 & limit\\
$[$Ar IV$]$ 4741    & 1.2825        & $<$3.0 & limit\\
H$\beta$            & 1.31473       & 32.4 & 1.1\\
$[$O III$]$ 4959    & 1.34111       & 49.3 & 1.6\\
$[$O III$]$ 5007    & 1.35409       & 159  & 1.4\\
\bf{Filter N6}\\
$[$N II$]$ 6548     & 1.77115       & 3.85  & 0.8\\
H$\alpha$           & 1.77488       & 116   & 1.2\\
$[$N II$]$ 6583     & 1.78002       & 7.4   & 1.2\\
$[$S II$]$ 6716     & 1.81663       & 6.7   & 0.3\\
$[$Ar III$]$ 7136   & 1.9296        & 2.8   & 0.3\\
\enddata
\tablecomments{Measured line fluxes.  
Columns are line ID, observed wavelength, 
lineflux in units of $10^{-16}$~\cgsflux, and uncertainty in the lineflux.
}
\end{deluxetable}

We assume a cosmology of $\Omega_m = 0.3$, $\Omega_{\Lambda} = 0.7$, and 
$H_0= 70$~\kms~Mpc$^{-1}$.  Solar abundances are taken from Table~1 of \citet{asplund}.  
The initial mass function (IMF) is \citet{chabrier03} unless noted.  

\section{Results}		\label{sec:results}

\begin{figure}
\figurenum{2}
\includegraphics[width=2.5in,height=6in,angle=90]{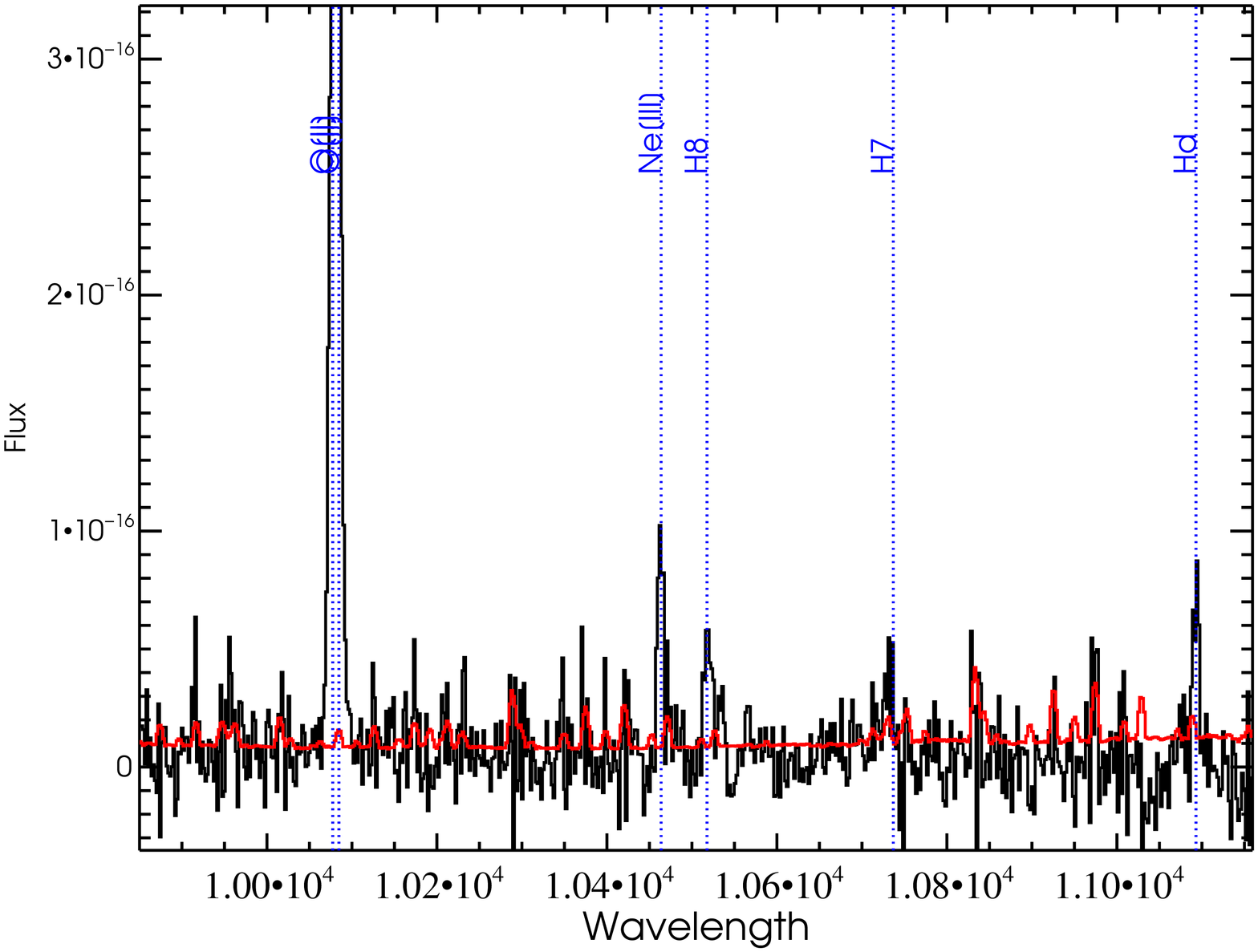}

\includegraphics[width=2.5in,height=6in,angle=90]{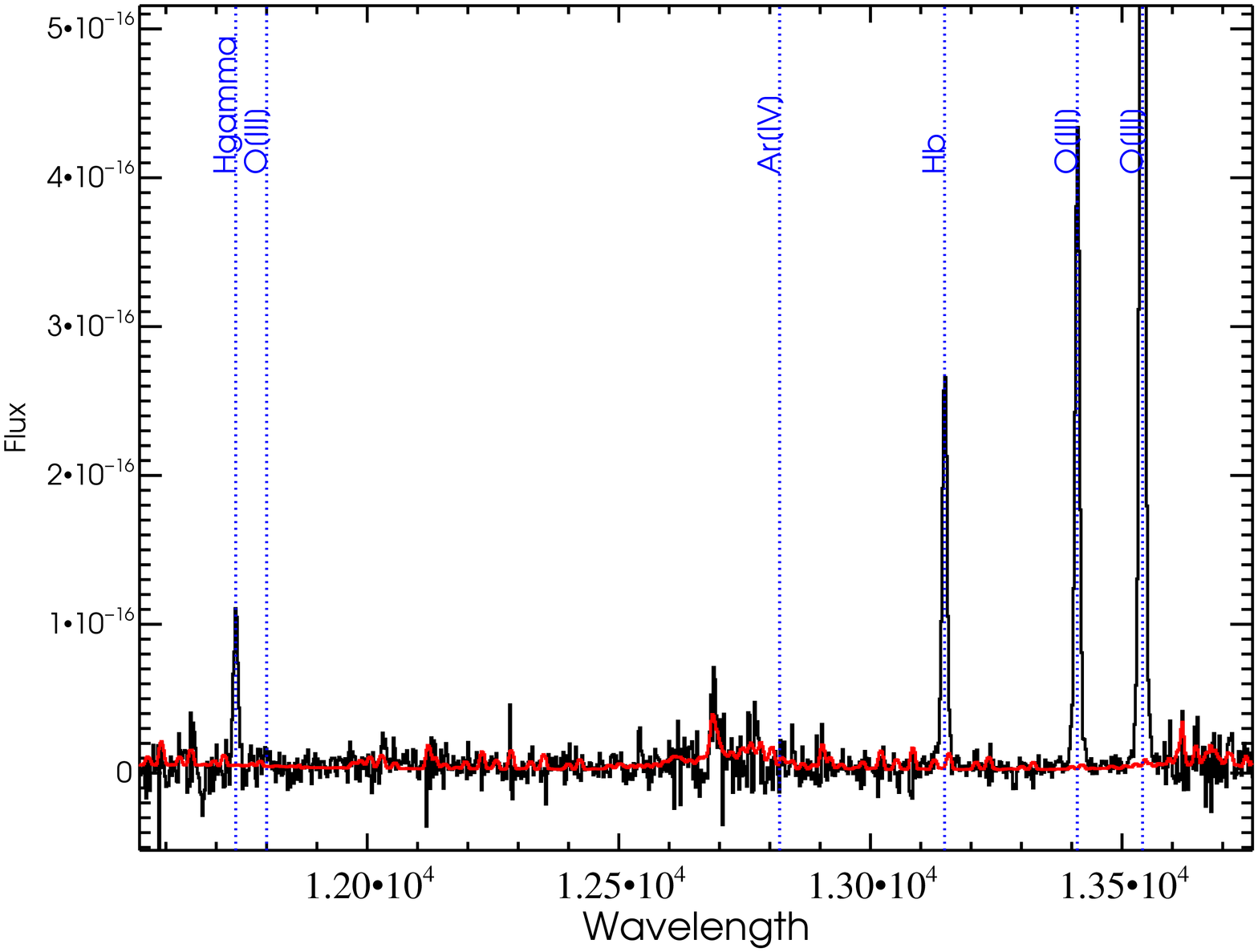}

\includegraphics[width=2.5in,height=6in,angle=90]{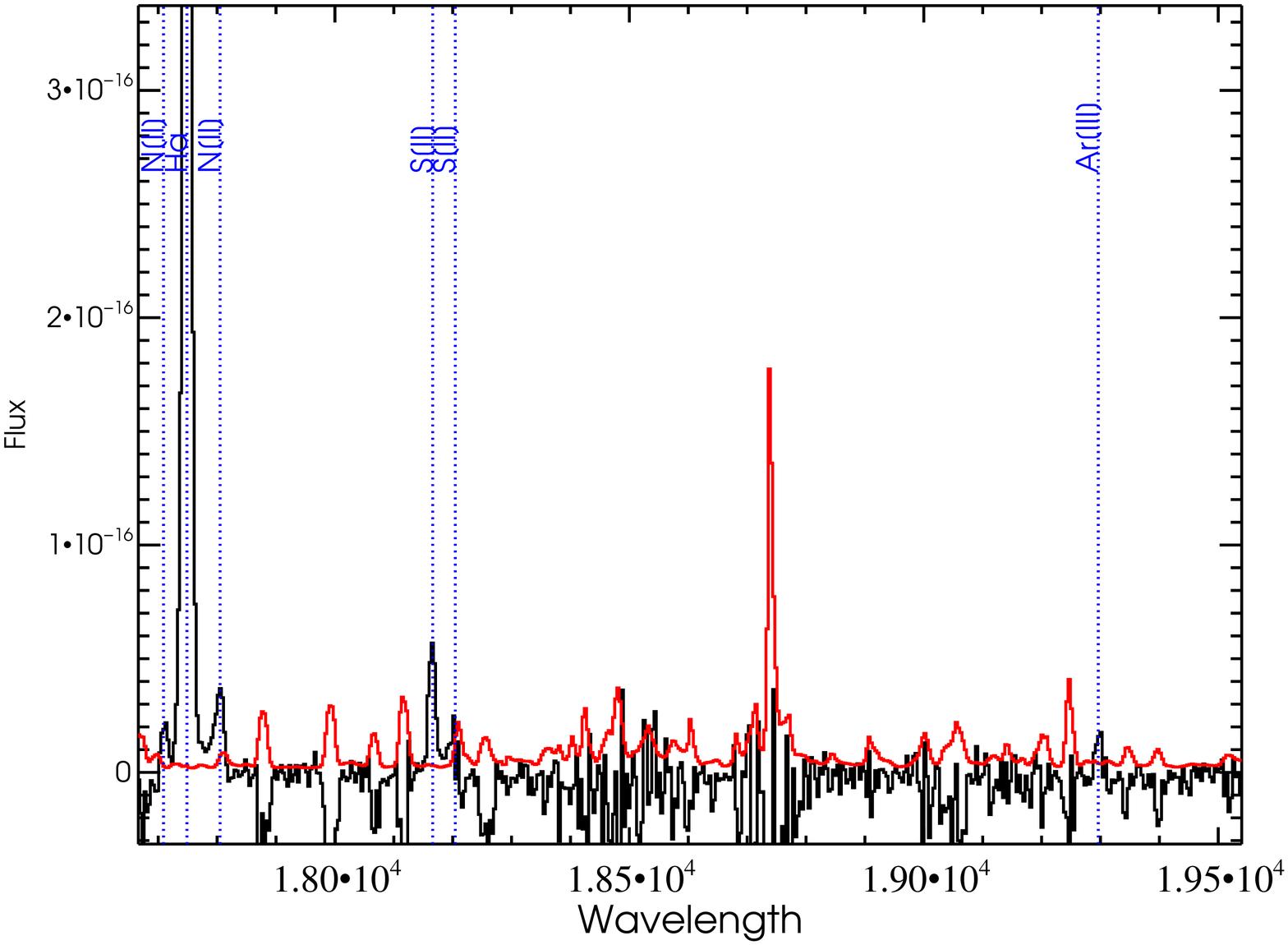}	
\figcaption{The NIRSPEC spectra.  The fluxed spectra are plotted in black, with the 
$1\sigma$ error spectrum in red.
The X-axis shows observed wavelength in Angstroms; the Y-axis shows observed flux in 
units of \cgsflux.
Blue labels mark detected emission lines, and lines for which we measure upper limits.
}
\label{fig:spectra}
\end{figure}

Figure~\ref{fig:spectra} plots the spectra.  
Several of the nebular emission lines are detected at very high signal-to-noise ratio (SNR); 
for example, we detect $\sim$50,000 net counts in the H$\beta$ line.  
Continuum is detected in filters NIRSPEC-1 and NIRSPEC-3 at SNR $\la 1$ per pixel.    
Continuum is not confidently detected near the lines of interest in NIRSPEC-6, 
which is unsurprising given the low atmospheric transmission at those wavelengths.  

\subsection{Extinction}\label{sec:extinction}

Based on SED fitting, \citet{wuyts} report low extinction values of E(B-V) $=$ 0.03 to 0.11, 
depending on metallicity and spatial position.  Given the potential degeneracies when fitting 
SEDs between extinction and age and star formation rate, a spectroscopic measurement of 
the extinction is desired.

The $\alpha$, $\beta$, $\gamma$, and $\delta$ transitions of the Balmer series are 
detected; H$\epsilon$ is formally undetected.  
The correction for stellar absorption,  $\sim$2\AA\ \citep{mccall}, is
negligible at these extremely high equivalent widths:
E$_r$ $=$ $430 \pm 70$~\AA\ and $100 \pm  20$~\AA\ for H$\beta$ and H$\gamma$.
 
H$\alpha$ appears in filter NIRSPEC-6;  H$\gamma$ and H$\beta$ in NIRSPEC-3;  
and H$\delta$ and H$\epsilon$ in NIRSPEC-1.  A line is detected at the position of H8, 
but is too strong to be H8 alone.  We suspect it is a blend of H8 with two He I lines, as 
in seen in the Orion Nebula \citep{orion}.

Because they appear in the same filter, 
the H$\beta$/H$\gamma$ and H$\delta$/H$\epsilon$ line ratios are free from relative fluxing errors.
We measure H$\beta$/H$\gamma$ $=2.39 \pm  0.37$, which for Case B recombination at 
T$=10^4$~K and n$_e= 100$~\cc\   
yields E(B-V) $=0.23 \pm 0.23$\footnote{Using the extinction law of \citet{ccm}.}
The H$\delta$/H$\epsilon$ ratio of $>$ 1.87 does not constrain the extinction given measurement errors.  
The best spectroscopic measure of extinction therefore comes from the H$\beta$/H$\gamma$ ratio, 
which for R$=$3.1, yields $A_v = 0.7 \pm 0.7$. 

The high end of this range is inconsistent with the blue color of the arc, as quantified 
by the SED fitting of \citet{wuyts}.  
Therefore, in the rest of the paper we will report results assuming first $A_v = 0.0$ and 
then $A_v = 0.7$.  
This considerable uncertainty in the extinction will limit how well we can measure 
spectral diagnostics that span a large range of wavelength, for example R$_{23}$.   
A more precise measurement of the extinction awaits simultaneous 
observations of H$\alpha$ and H$\beta$, or a longer integration on H$\gamma$ and H$\beta$.

\subsection{Tweaking the flux calibration} \label{sec:tweak}
As discussed in ~\S\ref{sec:methods}, there may be fluxing offsets among the three filters.  
We address this issue using the Balmer series.
Assuming an extinction of A$_v = 0.7$ as measured from the H$\beta$/H$\gamma$ ratio in NIRSPEC-3, 
we tweak the relative scalings of the NIRSPEC-1 and NIRSPEC-6 spectra to bring H$\alpha$/H$\beta$ 
and H$\delta$/H$\beta$ to the 
appropriate ratios for this extinction.
This increases the flux in  NIRSPEC-1 by a factor of 1.15 relative to NIRSPEC-3, 
and decreases the flux in NIRSPEC-6 by a factor of 0.61 relative to NIRSPEC-3.
These are reasonable corrections given the slit loss variation expected from seeing changes 
and atmospheric transparency issues in NIRSPEC-6. 
Fluxes reported in this paper reflect this scaling.

In Table~\ref{tab:fluxes} we report the line fluxes in filters NIRSPEC-1, -3, and -6.  
Line flux  ratios within a filter should be precise, limited by the uncertainty in 
line--fitting.  
When calculating a line flux ratio for lines spanning different filters, one should 
include a $10\%$ relative fluxing uncertainty. 

\subsection{Fraction of the galaxy covered} \label{sec:fraction}
To estimate the fraction of the galaxy covered by the slit, we use the
PANIC/Magellan images in J and K$_s$ of \citet{wuyts}, which had seeing of  
FWHM=0.57\arcsec\ and 0.51\arcsec.  
We estimate that $37 \pm 5\%$ of the total arc light falls into a slit 
positioned as in Figure~\ref{fig:finderchart}, neglecting slit losses.
We compute slit losses
for seeing of 0.8--0.9\arcsec\ and pointing errors of 0--0.25\arcsec.  
We conclude that $32\% \pm 4\%$ (random) $\pm 1\%$ (systematic) of the continuum light 
should have been captured by our NIRSPEC slit.

We now estimate this quantity in an alternate way. 
We measure the continuum adjacent to the 3737~\AA\ and H$\beta$ lines as 
$6.1 \times 10^{-18}$ and $6.0 \times 10^{-18}$~\cgsflam\ (using a robust average), 
with measurement errors of 10--20\%.
These continuua levels are $21\%$ of the continuum 
intensity in the spectral energy distribution 
in Figure~7 of \citet{wuyts}, which is the best fit to the  
u, B, g, r, I, z, J, H, Ks photometry of the whole arc.  

These two methods disagree by $38\%$.  We suspect that the continuum method is as fault, as 
the continuum is not well detected, and small DC offsets might plausibly be introduced 
in sky subtraction.   (Indeed, this uncertainty in the continuum level is why we chose to 
flux using AOV stars rather than multiplying line equivalent widths by f$_{\lambda}$ from 
broadband photometry as in \citealt{teplitz}.)  Therefore, we adopt the first method, and 
will multiply our NIRSPEC fluxes by the reciprocal, $3.12 \pm 0.4$, to convert to the 
expected line flux over the whole arc.

\subsection{Average Magnification}
\citet{wuyts} measure an average magnification of $u = 17.2 \pm 1.4$ for \arcname.  
Our current lensing model, based on ground--based photometry \citep{wuyts}, has 
insufficient spatial resolution to determine the magnification of each small knot.  
Thus, at present we cannot quote a reliable magnification for the portion of the arc 
that falls in our NIRSPEC slit.  It is likely that some of the bright knots 
are much more highly magnified than the arc on average.  For the time being 
we will adopt the average magnification, with these caveats.  Fortunately, except for 
the star formation rate derived from the Balmer lines, none of the results in this paper
depend on the magnification within the NIRSPEC slit.

\subsection{Redshift} \label{sec:z}
We fit the mean emission line redshift as $1.7037 \pm 0.0001$ using the 
NIST linelist,\footnote{http://www.pa.uky.edu/~peter/atomic/}, excluding 
[O II] 3727  because it is a partially resolved doublet. 
This compares perfectly with the redshift of rest-UV emission lines in 
our (not yet published) MagE/Magellan spectrum:  $1.70369 \pm 0.00006$. 
The MagE spectrum also yields a redshift for absorption lines in the interstellar medium:  
SiII 1260, Si II 1808, and Al II 1670, recommended to us by C.~Tremonti as isolated ISM lines,
yield an ISM redshift of $1.702671 \pm  0.0003$.  
Thus, the MagE spectra show that the interstellar medium is blueshifted 
from the nebular lines by $110 \pm 30$ \kms.
Our measured redshifts are 3 and 2$\sigma$ higher than the $1.7009 \pm 0.0008$ reported 
by Wuyts \etal\ (2010) using a combination of the [O II] 3727 and C III] 1909 emission lines 
and the Fe~II and Mg~II absorption lines.

\subsection{Velocity width} \label{sec:linewidths}
We extract arc lamp spectra using the same apertures as the science frames, and combine with the same
weighted average technique as for the science frames.
For each filter, we fit the instrumental linewidth versus wavelength relation with a linear fit, 
then interpolate the instrumental linewidth at the wavelength of each astronomical line of interest.
For example, at the observed wavelength of H$\alpha$, the arc lines have 
FWHM$=$13.6\AA\ compared to FWHM$=$15.3\AA\ observed for H$\alpha$; 
from this we infer $\sigma = 50 \pm 2$~\kms.
For H$\gamma$, H$\beta$, $\lambda$4959, and $\lambda$5007, we measure $\sigma$ = 76, 59, 49, 23~\kms.  
With H$\alpha$, this gives an average nebular line velocity dispersion
of $\sigma =  51$~\kms\ with an error in the mean of 9~\kms.  

We fit the [O II] 3726, 3729 doublet with two Gaussians, 
fixing the line centers using the NIST vacuum wavelengths and redshift from \S\ref{sec:z},
and forcing the two linewidths to be the same, 
while varying the continuum, line amplitudes, and the linewidth.  
This fitting finds that $\sigma = 51 \pm 1$~\kms\ for [O II].

Because the lensing morphology of \arcname\ is complex, we cannot yet
compute a meaningful effective radius, as needed to determine a dynamical mass 
from the velocity width.  
Upcoming HST observations should clarify this matter.

\subsection{Star formation rate}\label{sec:SFR}
\citet{wuyts}  estimated the SFR by fitting the broad-band optical and near--IR photometry, 
finding a SFR constraint of $<77 M_{\odot}~yr^{-1}$  for $40\%$ solar metallicity.
(For solar metallicity the SFR is lower).

We also estimate the star formation rate from mid--IR photometry. We 
measure the main arc and counter-image separately, finding MIPS/Spitzer 
24~\micron\ flux densities of $1040 \pm 153$ $\mu$Jy and $43 \pm 11$ 
$\mu$Jy respectively. Scaling by the average magnifications for the main 
arc and counter-image from the current lensing model 
\citep{wuyts}
yields de-lensed 24~\micron\ flux densities of $60 \pm 10$ $\mu$Jy 
for the main arc versus $21 \pm 6$ $\mu$Jy for the counter-image.  
The de-lensed 24~\micron\ flux density of the main arc corresponds to a star formation 
rate of $106 \pm 30 M_{\odot}~yr^{-1}$ using the prescription of 
\citet{rieke09}. 
The equivalent value from the counter-arc is $18 \pm 8$ M$_{\odot}~yr^{-1}$.

The significant difference between the de-lensed flux densities suggests that the source 
is strongly non-uniform; resolving this requires a more precise lensing 
model and a robust comparison of the observed image-plane data (both 
spectra and photometry) in the source plane. Until HST imaging of this 
source is available, the best comparison to the spectra presented here is 
the main arc, as the spectra sample a portion of that image. 

Thirdly, we  estimate the star formation rate from the NIRSPEC spectra, 
using the H$\beta$ line flux within the aperture,
assuming the extinction from H$\beta$/H$\gamma$, and the SFR(H$\alpha$) conversion of \citet{ken98araa}, 
modified for a \citet{chabrier03} IMF following Eqn.~10 of \citet{rieke09}:

\begin{equation} L(H\alpha)   = f(H\beta)                             (H\alpha/H\beta)   4\pi d_L^2     / u \end{equation}
\begin{equation}              = (3.2 \pm 0.1)\times 10^{-15}          (3.6 \pm 0.8)      4\pi d_L^2     / u \end{equation}
\begin{equation}              = (2.25 \pm 0.5)\times 10^{44} / u  \end{equation}
\begin{equation}              = (5.9  \pm 1.3)\times 10^{10} / u  L_{\odot} \end{equation}

\begin{equation} SFR(H\alpha) = 0.66 \times 7.9 \times 10^{-42} erg~s^{-1}  \end{equation}
\begin{equation}              = (1170 \pm 270 M_{\odot}~yr^{-1}) / u        \end{equation} 

A rough method of computing the total star formation rate is to divide by the 
average magnification from \citet{wuyts} and multiply by the fraction of the light in the 
image plane captured by the NIRSPEC slit (from \S\ref{sec:fraction}).  
This yields $SFR = 210 \pm 60 M_{\odot}~yr^{-1}$, which is extremely high.

The star formation rate derived from the Balmer lines is considerably higher than inferred from the 
broadband photometry or the 24~\micron\ flux.  The cause is not measurement error nor 
fluxing error.  (We have cross-checked the fluxing in multiple ways.)   
The Balmer lines are simply incredibly bright in these knots of \arcname, with 
H$\beta$ for example eight times brighter than in cB58.  

By spectroscopically targeting the brightest knots in \arcname, we may well be biased toward 
the regions with highest surface brightness and/or highest magnification.  The current
lensing map \citep{wuyts} lacks the high spatial resolution required to determine the 
magnification of each knot.   However, it does suggest that the magnification of this part of the 
arc is significantly higher than other parts of the arc.  
Thus, the rough method of computing the total SFR is probably wrong 
because it uses the average magnification of the whole arc, rather than the magnification of each 
pixel within the NIRSPEC slit.

A simpler method would be to measure the emission line flux over the entire arc, 
or at least all of the counter-arc, which is more compact and thus easier to map.  
Pending narrow-band HST imaging of H$\beta$ should provide this coverage for \arcname.

The three different methods give quite different answers for the star formation rate.
Only the brightest portion of the main arc   was observed with Nirspec;
this portion of the arc is possibly the 
brightest because of a conflation of a bright region within the source 
with a region of larger-than-average magnification, from which we would 
expect the spectral star formation rate to be larger than that derived 
from the average main arc 24~\micron\ flux density, as observed. A robust 
treatment of this comparison awaits a refined lensing model and 
reconstruction of source in the source plane.

Imaging is also biased toward regions of high surface brightness,  
and over--represents certain portions of the galaxy, but it 
does capture the entire image plane, and thus should deliver accurate quantities like 
average magnification, total magnitudes, and colors.

\subsection{Ionization parameter}

Figure 1 of \citet{KD02} illustrates the $\lambda$5007/$\lambda$3727 flux ratio 
as a diagnostic of the ionization parameter.  
For an O abundance of 20--40$\%$ solar (on the Asplund system), eqn.~12 of  
\citet{KD02}\footnote{Which uses the \citet{anders_grevesse} abundances; we convert to \citet{asplund}.}
yields an ionization parameter of $\log U = -2.73$ to $-2.85$ for $A_v=0.7$.  
For $A_v=0$, $\log U$ is higher by 0.1 dex.
The uncertainty due to extinction dominates over the flux uncertainty.

\subsection{Electron Density}  \label{sec:ne}
The two--component fit to the [O II] doublet in \S\ref{sec:linewidths} 
found a line flux ratio of f(3726/3729) $=0.893 \pm 0.024$, where 
this uncertainty is dominated by the uncertainty in deblending the doublet.
This ratio is density--dependent (c.f. Figure 5.8 of \citealt{OF06}.)
Using stsdas.analysis.temden in IRAF,\footnote{IRAF is distributed 
by the National Optical Astronomy Observatories,
which are operated by the Association of Universities for Research
in Astronomy, Inc., under cooperative agreement with the National
Science Foundation.}
our measurement corresponds to a 
tight measurement of the density:  $n_e = 235^{+28}_{-26}$ at $T_e = 10^4$~K, and
$n_e = 252^{+30}_{-28}$~\cc\  at $1.2\times 10^4$~K.

The ratio of the C III 1907/1909 doublet also constrains the density 
(\citealt{RubinFerland}, their Figure 2.)   
Fitting the doublet in our Mage/Magellan spectra (Rigby \etal\ in prep.) in the same 
way as the 3727 doublet, we measure a flux ratio of 
f(1907/1909) $= 2.16 \pm 0.3$.  
This ratio is unphysical by $1.7\sigma$, as the zero-density limit 
is 1.65 for pure C13, and 1.51 for pure C12 \citep{RubinFerland}. 
While deeper data should provide a better measurement of this line ratio, 
the current measurement of the C III] ratio is consistent with the low density regime of 
this diagnostic of $n_e \la 10^{3.5}$~\cc.

The doublet ratio of [S II]~6716, 6731~\AA\ also constrains the electron density 
(c.f.~Fig.~5.8 of \citealt{OF06}.)  
Unfortunately, the redder line is mostly lost to a skyline.  
Fitting in the same way as for the 3727 doublet, 
we measure a doublet ratio of $2.4 \pm 0.4$,  
which is unphysical by $2.4\sigma$ (the zero density limit is 1.43).
This unphysical result is likely due to the skyline contamination; we conclude that
in this case we cannot use the [S II] ratio to constrain the electron density.

To summarize, the [O II] 3727 doublet ratio provides a tight density constraint, 
which is consistent with the low density regime indicated by the C III] 1909 doublet. 
The [S II] doublet is contaminated by a skyline and provides no density constraint.

\subsection{Electron Temperature}\label{sec:Te}
The ratio [O III] ($\lambda$5007+$\lambda$4959)/$\lambda$4363
constrains the electron temperature with almost no $n_e$ dependence 
(c.f \citealt{izotov}; Figure~5.1 of \citealt{OF06}).  
The ratio in \arcname\ is $>139$ for $A_v=0.0$, and $>121$ for $A_v=0.7$.
Following \citet{izotov}, this corresponds to 
$T_e \le 1.14 \times 10^4$~K, 
$\le 1.20 \times 10^4$~K, and
$\le 1.26 \times 10^4$~K for A$_v=0.0$, $0.7$, and $1.4$.  

\subsection{Oxygen abundance} \label{sec:measure_O}
We constrain the oxygen abundance via the ``direct'' or ``T$_e$'' method
following \citet{izotov}, 
using the non-detection of [O III]~4363~\AA\ to constrain T$_e$, 
[O II]~3727~\AA\ and H$\beta$ to constrain (O$^+ / H^+$) and 
[O III] 4959, 5007, and H$\beta$ to constrain (O$^{2+} / H^{2+}$).  
Since He II 4686 is not detected, we can ignore the contribution of O$^{3+}$.
The result is  $12 + \log(O/H) > 8.21$ for $A_v=0.0$ and $>8.14$ for $A_v=0.7$; 
these are 0.48 and 0.55 dex lower than the solar value of $8.69\pm 0.05$ \citep{asplund}.

We compare this ``direct'' lower limit on the oxygen abundance to results from the 
bright line diagnostics.  We follow the calibrations for each diagnostic, then 
remove the relative offsets via the conversions of \citet{KE08}, 
so that all diagnostics are on the system of the N2 index of \citet{pettinipagel}.
The bright line results are as follows:

\begin{itemize}

\item The N2 index, $\log ([N II]/H\alpha) = -1.19 \pm 0.07$, 
yields $12+\log(O/H)=8.20 \pm 0.04$  by the third-order polynomial calibration of  \citet{pettinipagel}, 
and $8.22 \pm 0.04$ by their linear fit.  
Each of these two calibrations was reported to have a $1\sigma$ spread against T$_e$ of 0.18 dex 
\citep{pettinipagel}.  Reddening is irrelevant.

\item The N2 index calibration of \citet{D02} yields  $12+\log(O/H)=8.20 \pm 0.05$. 

\item The O3N2 index, $\log[(5007/H\beta)/(NII/H\alpha)]=1.88 \pm 0.07$, 
yields  $12+\log(O/H)= 8.16 \pm 0.02$   by the calibration of \citet{pettinipagel}, 
which has a $1\sigma$ spread of 0.14 dex.   Reddening is irrelevant.

\item The index $\log [N II 6584]/[O II 3727]$ is $-0.99 \pm 0.07$ for $A_v=0$ and 
 $-1.19 \pm 0.07$ for $A_v=0.7$.  The calibration of \citet{KD02} as modified by \citet{KE08}
produces a double--valued abundance:  
$12+\log(O/H)= 8.21$, $8.34 \pm 0.06$ for $A_v=0$; and 
$12+\log(O/H)= 8.21$, $8.22 \pm 0.06$ for $A_v=0.7$.

\item The Ne3O2 index, $\log(  [Ne III] \lambda 3869 / [O II] \lambda 3727) = -0.96 \pm 0.07$, 
yields  $12 + \log(O/H) = 8.19 \pm 0.08$ via the calibration of \citet{shiNe}.  
Reddening is irrelevant.  
This diagnostic does not appear in \citet{KE08}, so we cannot remove any calibration offsets.
\end{itemize}

The R$_{23}$ index, $\log[ (\lambda 3727 + \lambda 4959 + \lambda 5007)/H\beta]$, is
unfortunately double--valued as well.  
We measure $\log R_{23}$ as 0.94, 0.96, 0.99 for A$_v$=0.0, 0.7, 1.4, all with 
uncertainties of $\pm 0.02$ from the propagated flux uncertainty.
(This is similar to the value of 0.92 measured for cB58 by \citealt{teplitz}.)
Such a high R$_{23}$  requires a high ionization parameter 
($\log U \ge -2.87$ from Fig.~5 of \citealt{KD02}).  
The several R$_{23}$ calibrations methods  use various means to separate the 
``lower branch'' and ``upper branch'', typically [N II]/[O II] or [N II]/H$\alpha$ \citep{KE08}.  
Unfortunately, for \arcname\ these ratios are on the border between upper and lower branch.  
Thus, we compute the abundance for each branch:
\begin{itemize}

\item \citet{Z94} published an R$_{23}$ calibration for the upper branch only.   
We use it with caution
since \arcname\ is not clearly in either the upper or lower branch.  
For $A_v=0$, this calibration yields $12+\log(O/H)= 8.15 \pm 0.02$.
For $A_v=0.7$, it yields $8.14 \pm 0.02$.

\item The \citet{P05} method yields $12+\log(O/H)= 8.26 \pm 0.05$ and $8.07 \pm 0.08$ 
for the upper and lower branches for $A_v=0$, and 
$8.19$ and $8.22$ (same uncertainties) for $A_v=0.7$.

\item The \citet{KK04} method yields $12+\log(O/H)= 8.19$ and $ 8.16 \pm 0.02$ 
for the upper and lower branches and $A_v=0$, and 
$8.17 \pm 0.02$ for both branches at $A_v=0.7$. 
\end{itemize}

These inferred abundances are plotted in Figure~\ref{fig:Zs}.

\subsection{Abundances of other elements}\label{sec:otherelements}
Following \citet{izotov}, we constrain the abundance of N$^+$, Ne$^{2+}$, Ar$^{2+}$, and  Ar$^{3+}$ 
relative to H$^+$. 
Unlike O, the full suite of ionization states is not observed for these elements, 
so we must apply ionization correction factors to infer abundances.
\begin{itemize}

\item {\bf N:}  We derive $12 + \log(N/H)>6.53$ ($>6.42$)   for $A_V=0.0$ (0.7).  
Thus, N is depleted relative to the solar value by no more than 1.41 dex.
The uncertainty from extinction is larger than from flux uncertainties.

\item {\bf Ne:}  We derive  $12 + \log(Ne/H) > 7.33 \pm 0.09$, which is $0.60$ dex below solar.  

\item {\bf Ar:}  We derive  $12 + \log(Ar/H) > 5.54 \pm 0.18$, which is $0.86$ dex below solar. 
\end{itemize}

Though we have measured lower limits on all abundances, we can tightly constrain the key 
abundance ratios, which depend only weakly on $T_e$:
\begin{itemize}

\item {\bf N/O}:  For the maximum allowed $T_e$,  $\log(N/O)=-1.70\pm 0.02$.
If the $\lambda$4363 flux is one-third the upper limit, then $T_e$ is $70\%$ of the maximum, 
and  $\log(N/O)=-1.89\pm 0.04$.
These ratios are $0.84\pm0.02$ to $1.03\pm 0.04$ dex below the solar abundance ratio.  

\item {\bf Ne/O}: For the maximum allowed $T_e$,  
$\log(Ne/O)=-0.89 \pm 0.09$ for $A_v=0$, which is 0.13 dex below the solar ratio.
For  $A_v=0.7$ the ratio is $\log(Ne/O)=-0.72 \pm 0.09$, which is 0.04 dex above solar.  
At $70\%$ of the maximum permitted $T_e$, 
the constraints are from 0.03 dex below to 0.14 dex above solar.  
Thus, no matter the actual $T_e$, the Ne/O ratio must be solar--like.

\item {\bf Ar/O}:  For the maximum allowed $T_e$, $\log(Ar/O)=-0.89\pm 0.18$, 
which is 0.13 dex below the solar value.
For  $70\%$ of the maximum permitted $T_e$, 
$\log(Ar/O)=-0.62\pm 0.18$, which is 0.14 dex above solar.
\end{itemize}

Thus, \arcname\ has an abundance pattern in which the 
O abundance is at least $29\%$ of solar, 
the Ne/O and Ar/O ratios are solar--like, and 
the N/O ratio is less than $15\%$ of the solar ratio.

\subsection{Joint constraints on physical conditions from photoionization models}
As a cross-check on these diagnostics, we run spectral synthesis and photoionization models.  
The UV spectra come from Starburst 99 (v5.1, web version; \citealt{s99a} and \citealt{s99b}), 
assuming continuous star formation, 
an upper mass cutoff of 100~\Msol, the default IMF parameters, the Padova AGB models, 
and with stellar metallicity set at $40\%$ of solar.  
We feed these UV spectra to the photoionization code Cloudy version c08.00 \citep{cloudy}, 
running a grid of models of given electron density and starburst age, 
and in each model optimizing the metallicity and ionization parameter, 
as constrained by the measured line intensities relative to H$\beta$, assuming a 
foreground screen extinction of $A_v=0.7$, the measured uncertainties for lines 
contiguous with H$\beta$, and 
 $20\%$ relative fluxing errors for noncontiguous lines.     
The abundance pattern is ``H II'' region.  
Our grid of electron density covers $\pm 1\sigma$ of the value measured in \S\ref{sec:ne}.

Cloudy computes an average electron temperature of $T_e = 1.19$ to $1.24 \times 10^4$~K, 
which is consistent with the value inferred in \S\ref{sec:Te}.
Cloudy converges on an  oxygen abundance in the range $12 + \log(O/H) = 8.07$ to 8.13, 
and an ionization parameter in the range $\log U = -2.57$ to $-2.42$.  
Figure~\ref{fig:cloudy} illustrates these constraints.  
Thus, Cloudy converges on an O abundance that is 0.1 dex lower than via the $T_e$ method, 
and a $\log U$ that is higher by 0.2--0.3 dex than derived from the \citet{KD02} method.  

The source of this discrepancy may lie with different assumptions of abundance pattern, 
filling factor, atomic data, or ionizing stellar spectra compared to \citet{KD02}.  
Resolving this moderate discrepancy is outside the scope of this paper at this time, but it
provides a caution that one must be careful, when comparing derived physical properties of galaxies, 
to use the same methodology.  
In this paper, we will adopt the $\log U$ derived via the \citet{KD02} method, so that we 
may compare to other lensed galaxies in the literature.  

Dust grains were not included in the Cloudy models.
As a test, we added dust grains with the same metallicity 
and abundance pattern as the gas, which changed the inferred 
metallicity and ionization parameter by only $2\%$ and $1\%$.

\section{Discussion}			\label{sec:discussion}

Since this is the  first investigation of the rest-frame 
optical lines in \arcname, only the brightest ``knots'' were targeted, representing 
only $32\% \pm 4\%$ (random) $\pm 1\%$ (systematic) of the 
total light of the arc (\S\ref{sec:fraction}).  
\arcname\ was selected as the brightest and one of the largest 
lensed sources in a large survey; despite its exceptional appearance,
SED and lensing modeling show it to be a fairly typical object \citep{wuyts}.
As such, it is likely that the high surface brightness portions of the source 
galaxy coincide with high magnification, and that this coincidence makes \arcname\
such a spectacular example of strong lensing.  This supposition would explain the 
differing star formation rates derived from 24$\mu$m imaging of the main arc and 
counter-image, as noted in \S3.7.  
Moreover, because the lensed source straddles the lensing
tangential caustic (c.f. Figure 6 in \citealt{wuyts}),  the lens model
suggests that only some regions of the galaxy are represented in the
giant arc, namely the core, and one of two apparent spiral arms, and
may be highly magnified. Thus the giant arc is likely not a fair
representation of the source galaxy as a whole.

Since superb spectra can be 
quickly obtained for these bright knots, it is their physical conditions that we probe 
in this paper, with a caveat that these may not be ``ordinary'' regions of this galaxy.
Once the arc has been fully mapped in these emission lines, 
and a high-resolution lensing map has been derived from HST imaging,  we should understand how
much the physical conditions vary across the source plane, and locate the brightest 
knots in the source plane.  It will then be possible to fully contextualize our results 
in terms of the range of physical conditions across the spatial extent of \arcname.

For now, we assume that the knots for which we have spectra are representative.  
Since these are highest-quality rest-frame optical spectra ever obtained for a  z$=$2 galaxy, 
and yield precise measurements of the physical conditions, we now consider these measurements 
in the context of the literature, and the conditions under which stars form in the distant universe.  

\subsection{Extinction}
The following E(B-V) extinctions have been measured from Balmer decrements in 
lensed blue galaxies: 
$0.27$ for cB58 \citep{teplitz}; 
$0.28 \pm 0.04$ for the Clone \citep{hainline}; 
$0.45 \pm 0.04$ for the Cosmic Horseshoe \citep{hainline}; 
$0.59 \pm 0.08$ for J0900 \citep{bian}; 
$0.67 \pm 0.21$ for the 8 o'clock arc \citep{finkelstein}.
Our measurement of E(B-V) $=0.23 \pm 0.23$ for \arcname\ is consistent with the 
lower part of this range, but the uncertainty is large, stemming mostly from the 
flux uncertainty in H$\gamma$.  A spectrograph that obtains H$\alpha$ and H$\beta$
simultaneously, for example FIRE on Magellan or LUCIFER on LBT, would provide 
a very precise measurement of the Balmer decrement in this galaxy, allowing a detailed 
comparison of the relative extinctions suffered by the gas and the stars.

\subsection{The reliability of O abundance diagnostics at z$=$2}
For \arcname, the non-detection of [O III] 4363~\AA\ sets a strict constraint on the
oxygen abundance via the ``direct'' T$_e$ method: it must be no more than 0.55 dex (0.48 dex) 
below the solar value for $A_v=0.7$ ($A_v=0$), or $28\%$ and $33\%$ as percentages of solar.  
Neon provides a cross-check on this abundance measurement.  Since both Ne and O are 
alpha--group elements, their ratio should be solar--like, 
and indeed our measured ratio is solar with an uncertainty of $\pm 0.14$ dex.  
We believe this to be the first time the Ne/O ratio has been measured at $z\sim2$.  

We now accept the $T_e$--method's oxygen abundance measurement, 
and assess the reliability of the ``bright line'' diagnostics.  

These diagnostics are empirically calibrated at z$=$0 against the $T_e$ method, or calibrated with 
photoionization models.  As such, they implicitly assume densities and ionization parameters
typical of nearby galaxies, which may not be appropriate for high--redshift galaxies, 
for reasons of luminosity bias and evolution in the galaxy population.  
In addition, these diagnostics have significant relative offsets (up to 0.7 dex) at z$=$0 whose 
origins are not understood and which must be empirically calibrated away using 
SDSS galaxies \citep{KE08}; it is not clear that for rapidly star--forming z$=$2 galaxies 
these offsets will be the same.

Thus, it is important to test these bright--line diagnostics \textit{in situ} at z$=$2.
Figure~\ref{fig:Zs} compares the O abundances inferred in \S\ref{sec:measure_O} for the 
bright--line diagnostics in \arcname.  For ease of comparison, we plot $A_v$$=$0 and $A_v$$=$0.7 
separately.  
We divide the calibrations into five categories:  [N II]/[O II], Ne3O2, N2, O3N2, and R$_{23}$ methods.

The [N II]/[O II] diagnostic \citep{KD02} is consistent with the $T_e$ method, 
in part because its double-valued results and the large uncertainty covers 0.3 dex of parameter space.  
Because this index has a large extinction correction, we cannot adequately test it until
the extinction of \arcname\ is more precisely measured.

The Ne3O2 diagnostic is consistent with the $T_e$ method, especially for 
$A_V>0$.  

The N2 calibrations of  \citet{pettinipagel} and \citet{D02} are consistent with the lower limit from the 
$T_e$ method, especially for $A_V>0$.  This test is especially importance, since this diagnostic 
is the one used to measure the evolution in the mass--metallicity relation from z$=$0 
to z$=$2 \citep{erb06} and z$=$3 \citep{mannucci,maiolino08}. 

The O3N2 diagnostic \citep{pettinipagel} yields an abundance that is $2.5\sigma$ 
below the $T_e$ lower limit for $A_v=0$, and is thus inconsistent with the $T_e$ limit.
However, moderate extinction ($A_v=0.7$) makes the results consistent.  
The O3N2 result is lower than the abundances from the N2 and [N II]/[O II] indices.  

The  R$_{23}$ index has a large extinction correction.  As such, it is difficult to assess the performance
of this diagnostic given the current uncertainty in the extinction of \arcname.  
For $A_v=0$ only the upper branch of \citet{P05} is consistent with the $T_e$ result; 
for $A_V=0.7$ the \citet{P05} method is also consistent.  
The \citet{Z94} method produces an inconsistent result, but this is unsurprising since it is only valid for 
high metallicity.

This is the first test \textit{in situ} at z$=$2 of the bright line abundance diagnostics for 
a star--forming galaxy of typical metallicity.
To summarize, the R23 and [N II]/[O II] methods are not inconsistent with the $T_e$ method so long as 
the extinction exceeds zero; the definitive test must await a more precise extinction measurement, since 
both diagnostics have large reddening corrections.  
The Ne3O2 diagnostic is consistent with the $T_e$ method, though the errorbars 
are larger than for N2 and O3N2, since [Ne III] is not a terribly bright line.
The N2 and and O3N2 methods are most constraining given the current extinction measurement; of these, 
the N2 results are entirely consistent with the $T_e$ method, and the O3N2 diagnostic is consistent 
only given significant but plausible extinction.

The only other published $T_e$ detection at z$=$2 of which we are aware is in Lens22.3, a low--mass, 
low--metallicity galaxy behind Abell 1689 \citep{yuankewley}.  
For Lens22.3, the $T_e$ method yields $12+\log(O/H) = 7.5 \pm 0.1$ 
for zero extinction and $7.3 \pm 0.1$ for high extinction. 
We apply the  \citet{pettinipagel} N2 and O3N2 calibrations to the 
published line fluxes of \citet{yuankewley} (N II is not detected).
From N2 $<$ -2.30 comes an abundance of $12+\log(O/H) < 7.47$  ($7.59$) using the 
third-order (linear) N2 calibration.
The O3N2 index,  $\log O3N2 > 3.11$, yields $12+\log(O/H) < 7.73$ on the original \citet{pettinipagel} 
scale, which is outside the range that can be converted to the N2 frame \citep{KE08}.  
These N2 and O3N2 measurements are consistent with the $T_e$ result.  
However, all these diagnostics are highly inconsistent with the R23 result of 
$12 + \log(O/H) = 8.0$--8.3  \citep{yuankewley}.

\citet{yuankewley} noted the discrepancy between the R23 and $T_e$ results, and attributed the fault 
to the $T_e$ method.  Given the consistency between the N2, O3N2, and $T_e$ methods, we suggest
that it may instead be R23 that is failing in Lens22.3. 

\subsection{The reliability of O abundance diagnostics in other z$=$2 lensed galaxies}

Only in \arcname\ and Lens22.3 are the spectra sufficient to measure abundances via the 
$T_e$ method.  However, the bright line diagnostics have been used for a number of galaxies.  
A re-analysis of the mass-metallicity relation at $z\ge2$ is outside the scope of this paper.  
However, it is appropriate at this juncture to re-examine the N2 and O3N2 diagnostics in 
light of the results above.
In Figure~\ref{fig:massZ} we plot the mass-metallicity relation at z$=$2, using
literature results for lensed and unlensed galaxies, and our results for \arcname.  
For fair comparison, we convert stellar masses to the \citet{chabrier03} IMF, take line fluxes 
from the literature and apply the N2 and O3N2 calibrations of \citet{pettinipagel}, 
brought to the N2 system via the conversion\footnote{This conversion is modest 
($<0.05$ change in $12+\log(O/H)$) at these metallicities.} of \citet{KE08}.

We plot four lensed galaxies in Figure~\ref{fig:massZ}:  
cB58 \citep{teplitz}, J0900+2234  \citep{bian}, the 8 o'clock arc \citep{finkelstein}, 
and \arcname\ (this work).    These four galaxies span a range of 400 in stellar mass, 
extending below the mass range probed by stacked samples \citep{erb06}.  
In all four cases, the O3N2--derived abundances are systematically lower than those
derived from N2, strikingly so in the case of the 8 o'clock arc.
(Recall that the relative offset between these indices, as measured in SDSS at z$=$0, 
has already been removed.)
These offsets were noted for each galaxy in their respective papers, but were 
dismissed as less than the calibration dispersion of the diagnostics.  
The Clone and Horseshoe lack stellar mass measurements to plot them on Figure~\ref{fig:massZ}, 
but comparison of their N2 and O3N2--derived abundances shows the same trend:  the O3N2 index 
is low by 0.16 and 0.07 dex. 

It is clear from a consideration of all six lensed galaxies that
the O3N2 and N2 indices are systematically offset at z$=$2.  While this offset is 
small for the galaxies of intermediate metallicity, it is large
for the 8 o'clock arc at the high metallicity end.  

This behavior is not surprising when we consider that O3N2 has 5007/H$\beta$ in the numerator,
and that what makes z$=$2 galaxies stand out in the BPT diagram \citet{bpt}
is their high 5007/H$\beta$ compared to SDSS galaxies.
Thus, we suggest that high ionization conditions in z$=$2 galaxies cause the O3N2
diagnostic to function poorly at these redshifts.

\subsection{Lensed galaxies and the z$=$2 mass-metallicity relation}

\arcname\ has a stellar mass of 
$\log (M_*) = 10.0 \pm 0.1$~\Msol\ \citep{wuyts}; both the main arc and the counter-arc 
give consistent results.  This is $0.94 \pm 0.14$ dex lower than 
the Schechter function parameter $M^*_{star}$ at z$=$2.0 \citep{danilo}, 
and lower than most of the Lyman break galaxies in \citet{erb06}.  
The low measured velocity dispersion qualitatively supports a low mass.  
In the mass-metallicity plane, 
\arcname\ lies significantly, $0.17 \pm 0.04$ dex, below the z$=$2 
relation of \citet{erb06}.  This offset is too large to be measurement 
error, and so we attribute it instead to a real abundance difference: \arcname\ is 
metal--poor by about $50\%$ for its stellar mass and redshift, compared to \citet{erb06}.

J0900+2234 shows this effect to an even greater degree \citep{bian}, since it 
has almost the same N2--derived oxygen abundance and a slightly higher stellar mass.

On the high-mass end, the N2--derived oxygen abundance of the 8 o'clock arc is quite 
consistent with the high--mass side of \citet{erb06}, perhaps 0.05--0.08 dex low.  
By contrast, at the very low--mass end, cB8 lies far off the relation, having a very low stellar mass
but a high oxygen abundance, more typical of the mass-metallicity relation at z$=$0 than at z$=$2--3.  
Thus, even from a modest sample of four lensed galaxies, we are already 
beginning to probe the intrinsic spread in the mass--metallicity relation at z$=$2.

\subsection{Abundance pattern}

In \S\ref{sec:otherelements} we constrained the N/O ratio as 
$\log(N/O) \le -1.70$.
For plausible values of $T_e$ the ratio can be lower by 0.2 dex.  
These abundances are startlingly close to the values measured for cB58:
$\log(N/O)=-1.76 \pm 0.2$ dex \citep{pettini02} and $12 + \log(O/H)=8.26$ \citep{teplitz}.\footnote{\citet{teplitz} 
infer an N/O ratio that is higher by 0.5 dex, a discrepancy discussed by \citet{pettini02}.}
Even though \arcname 's stellar mass is $18$ times larger than cB58's,  and its redshift lower, 
star formation in these two galaxies has produced very similar ratios of N, O, and H.  
UV spectroscopy for \arcname, as in \citet{pettini02} for cB58, should allow element-by-element 
comparison of abundance ratios.

For now, we concentrate on the N/O ratio.
\arcname\ lies near the intersection of the primary and secondary N production lines.   
In other words, it lies right on the trend for secondary N production, and is somewhat below 
the primary plateau of $\log(N/O)\sim -1.5$.    
As such, its N/O ratio is at the low end of what has been measured for H II regions in spiral galaxies 
of comparable O abundance, and is typical of dwarf galaxies \citep{vanzee}.
Analytic models such as  \citet{henry} combine both N production mechanisms to produce smooth 
curves of N/O versus O.  \arcname\ lies on these curves near the ``knee'',  where
the N/O ratio rapidly transitions from being
independent of the oxygen abundance (``primary'' production), to being highly dependent on 
the O abundance (``secondary'' production).   
Comparing \arcname\ and cB58 to the numerical models of \citet{henry} (their Figure 3b) suggests
that both galaxies have relatively high star formation efficiencies.
With a sample of two galaxies, it is premature to draw  conclusions about how galaxies build up 
their nitrogen,  but the strict N/O and O measurements we have made bode well for the future, if such work
can be repeated for a larger sample.

\subsection{Ionization parameter}
Ionization parameters have been reported for a few other lensed galaxies, derived from the 
5007/3727 diagnostic.  Unfortunately, this diagnostic has a strong dependence on O abundance, 
so this must be input to the diagnostic.    %
We use the 5007/3727 line fluxes reported by \citet{hainline} for the Horseshoe and the Clone, 
assume $40\%$ solar metallicity (Asplund system), as inferred for both from the N2 index.  
Converting to \citet{anders_grevesse} abundances and using the calibration of \citet{KD02},
this corresponds to an ionization parameter of 
$\log U = -2.8$ for the Horseshoe and $-2.7$ for the Clone, without correcting for extinction. 
This is consistent within 0.1 dex with the values reported by \citet{hainline}.
\citet{teplitz} did not calculate an ionization parameter for cB58, so we take their line fluxes and 
the O abundance from \citet{pettini02}, and find $\log U=-2.85$ for cB58 without correcting for 
extinction.  Extinction dominates the uncertainty on this measurement, 
and acts to lower the ionization parameter.

These ionization parameter measurements are entirely consistent with the value measured 
for \arcname\ of -2.73 to -2.85 (for A$_v$ $=$ 0.7; 0.1 dex higher for A$_v$ $=$ 0).  
Thus, we conclude that the four lensed galaxies examined to date with this diagnostic 
all have very similar ionization parameters of about $\log U \sim -2.7$.

We now compare to local galaxies.   The sample of \citet{Kewley_IR} would be 
ideal for comparison, since these are IR--luminous galaxies at z$\sim$0, but 
unfortunately their spectroscopy did not extend blueward to  3727~\AA.  
Instead, we turn to two papers which measured ionization parameters
for 
65 SINGS galaxies \citep{moustakas_10}, 412 star--forming galaxies, and 120,000 galaxies from 
the Sloan Digital Sky Survey \citep{moustakas_06}, using the same methodology 
of calculating $\log U$ as we do.
The SINGS galaxies have a median and mode $\log U$
of $-3.0$.  The SDSS and 412 star--forming galaxies (figure 12 of \citealt{moustakas_06}) have a
narrow range of $-3.2 < \log U < -2.9$.  
The prototypical starburst galaxy M82 has $\log U = -2.9 \pm 0.07$.

Thus, the four lensed galaxies with measured ionization parameters all 
have the same measured value, within the uncertainties.  This value, $-2.7$ to $-2.8$, 
is roughly twice as high as the median for samples of lower--luminosity
nearby galaxies, and is 30--$60\%$ higher than for M82.  
Thus, the ionization parameters of these four z$\sim$2 star-forming galaxies 
are somewhat higher than local galaxies of much lower luminosity.

\subsection{Electron density}

In \S\ref{sec:ne} we derived an electron density of $n_e = 252^{+30}_{-28}$~\cc\  at 
$1.2\times 10^4$~K from the [O II] 3726, 3729 doublet ratio.  
This is currently the most precise such measurement at such high redshift.  
We now compare to density constraints from the literature for other lensed galaxies.

\citet{hainline} measured the [S II] $\lambda$6717, $\lambda$6732 flux ratio
in two lensed galaxies.
For the Clone, the result is $0.9 \pm 0.1$ by extracting the fluxes in each of two apertures 
and summing, versus $0.75 \pm 0.25$ by extracting all at once.
For the Horseshoe, the result from the summed aperture is $1.0 \pm 0.35$.
Using IRAF's temden at $T_e=10^4$~K, these ratios indicate densities of 
$900^{+500}_{-300}$~\cc\ and $1700^{+11000}_{-1100}$~\cc\ for the Clone, 
and $600^{+2400}_{-500}$~\cc\ for the Horseshoe.
\citet{bian} measured a [S II] flux ratio of $0.86 \pm 0.2$ for the sum of both apertures in J0900, 
which yields a density of $1100^{+1700}_{-600}$~\cc.  
In the rest-UV, \citet{horseshoe_quider} measure a  C III] flux ratio of 
$f(1906)/f(1908)=1.1 \pm 0.2$ for the Horseshoe, corresponding to a density range of 5000--25000~\cc, 
which is inconsistent with the density range measured by \citet{hainline}.

Thus, these literature measurements of electron density are in the range 600--5000~\cc, albeit 
with large uncertainties.  The low precision of these measurements indicates a clear need for 
deeper spectra to better measure electron density.  Nevertheless, the current 
measurements for the Clone, Horseshoe, and J0900 favor high electron densities, much higher
than the precise value we measure for \arcname.  Thus, at present 
it is not clear what is a typical electron density for a star forming galaxy at 
these epochs.  Additional high--quality measurements are urgently needed.

\subsection{Location in the BPT diagram}

The ``BPT'' diagnostic diagram of \citet{bpt} is commonly used to characterize 
the ionization conditions in galaxies.  For \arcname, we measure line ratios of 
log ([N II] 6583 / H$\alpha$) $=  -1.18 \pm 0.07$,  and 
log (5007/H$\beta$)     $=  0.69 \pm 0.02$ for Av=0 and 
0.13 less than that for $A_v = 0.7$.
In Figure~\ref{fig:BPT} we plot \arcname\ on the BPT diagram; 
its high OIII/H$\beta$ and extremely low NII/H$\alpha$ place it 
in the upper left quadrant, close to the maximal starburst line.   

This is an exceptional position compared to the z$=$0 SDSS, which has only 5 galaxies 
in that region of the BPT diagram. 
However, z$=$0 IR--luminous galaxies do occupy that space:  \citet{Kewley_IR} have
9 galaxies with NII/H$\alpha < -1$.  Since extreme star formation is rare in the local 
universe, these luminous star--forming galaxies may be a better basis for comparison than SDSS.  

It has previously been noted that z$\ga$1 galaxies tend to be offset toward higher 5007/H$\beta$
ratios in the BPT diagram \citep{shapley05,erb06,kriek07}.
\citet{BPC} proposed that this is caused by an elevated ionization parameter at 
higher redshift, and enumerated the following possible underlying causes:
a top--heavy initial mass function; higher electron densities; a higher volume filling factor; 
or a higher escape fraction of UV photons.
   
Figure~\ref{fig:BPT} shows that of the lensed galaxies, 
J0900 and the Clone show an offset similar to that of the \citet{erb06} stacked galaxies, 
while the 8 o'clock arc is considerably higher, as discussed  by \citet{finkelstein}.  
By contrast, \arcname\ and cB58 are not as offset -- they lie between the \citet{erb06} 
points and the z$=$0 IR--luminous galaxies.\footnote{The Horseshoe is not plotted 
because H$\beta$ was contaminated by a skyline \citep{hainline}.  
CB58 was not plotted because [S II] was not observed.}

\arcname\ has a electron density ($n_e = 235^{+28}_{-26}$~\cc) that is lower than 
the best-fitting densities for the Clone, Horseshoe, and J0900, though those other measurements 
have very large uncertainties.
Its measured ionization parameter ($2.9 \pm 0.17$ for $A_v=0.7$) 
is entirely consistent with measurements of the same diagnostic in the Horseshoe, Clone, and cB58.  
Thus, in two ways \arcname\ contradicts the picture of \citet{BPC} for BPT behavior at z$=$2:  
First, though its ionization parameter is the same as the other z$=$2 galaxies, 
it is not as offset in the BPT diagram; 
second, the electron density is not high, as has been suggested to explain offsets in the BPT diagram.

Of the IR--luminous sample of \citet{Kewley_IR} with NII/H$\alpha < -1$, the median electron density is $\sim 100$~\cc.  
This is another case in which offsets in the BPT diagram do not appear to be caused by high density.

At present, too few z$=$2 galaxies have the high--quality spectroscopy necessary to fully map their behavior
in the BPT diagram, and link offsets back to  evolution in physical conditions.  
This will presumably change as new lensed galaxies are pursued, and as 
new multi-object near-IR spectrographs push down the luminosity function of the non-lensed 
galaxy population.
That said, \arcname\ demonstrates that offsets in the BPT diagram are not driven by 
higher ionization parameters and densities, as has been suggested.
It also demonstrates that star--forming z$=$2 galaxies may have quite similar 
ionization parameters and densities to z$=$0 galaxies.

Last in our discussion of the BPT diagram, we note that \arcname's extreme location 
places it as far as possible from the z$=$0 AGN locus.  Thus, it is unlikely that the nebular
emission of this galaxy is dominated by an AGN, though we have not  ruled out a low--luminosity AGN
(c.f.~\citealt{greene}).  Upcoming observations with the Chandra X-ray Observatory should test 
this.

\section{Conclusions}			\label{sec:conclusions}

The spectra published here total 1.3~hr of integration time, divided into 
20, 20, and 40 minute integrations over three spectral regions.
These spectra demonstrate  the power
of gravitational lensing to explore the physical conditions of star formation at the epoch 
when most of the Universe's stars formed.

For the second time in any galaxy at $z\sim2$, and for the first in an average--metallicity galaxy, 
we tightly constrain the O abundance using the ``direct'' $T_e$ method, thanks to the 
very constraining non--detection of [O III]~$\lambda$4363.  We use this constraint to test the 
bright--line diagnostics of oxygen abundance, which are the easiest to measure, 
but are empirically calibrated at z$=$0 and thus incorporate assumptions about density and 
ionization parameter that  may well be wrong at z$\sim$2.  

We find that the O abundance inferred 
from the  N2 index (the ratio of [N II]/$H\alpha$) agrees closely with the $T_e$ method 
and has a small uncertainty.  Ne3O2 also performs well, albeit with a larger uncertainty since [Ne III] 
is not a bright line.  N2O3 and R23 depend so strongly on the extinction that we cannot definitively 
assess their performance in \arcname, though they appear to work for the current best estimate of the 
extinction, $A_v = 0.7$.    This is especially interesting because R23  fails in Lens22.3
\citep{yuankewley}, the only published example of an  [O III]~4363 detection at z$\sim$2.

The O3N2 diagnostic is on the border of disagreeing with $T_e$ in \arcname, depending on extinction.  
Comparing to the N2 index, O3N2 predicts systematically lower abundances in five z$\sim$2 lensed galaxies, 
dramatically so at near--solar metallicity.  We suggest that O3N2 does not work in z$=$2 galaxies, 
perhaps due to different physical conditions compared to z$=$0 where this diagnostic is calibrated. 

After all, O3N2 is effectively a location in the BPT diagram, and it has been shown, via small samples
of lensed galaxies and stacked samples of unlensed galaxies, that z$=$2 galaxies are offset in the
BPT diagram, with higher 5007/H$\beta$, than the cloud of SDSS galaxies at z$=$0.  
Indeed, \arcname\ is offset in the BPT diagram compared to the SDSS and \citet{Kewley_IR} galaxies, 
as offset as cB58 is, but not as offset as other z$=$2 galaxies.  

This is particularly interesting given that the measured ionization parameter for \arcname\ 
is the same as has been measured in the Horseshoe, the Clone, and cB58: $\log U \sim -2.7$ to $-2.8$, 
which is  30--60$\%$ higher than in M82.

We also tightly constrain the electron density: $n_e = 252^{+30}_{-28}$~\cc\  at $1.2\times 10^4$~K, 
which is not terribly high compared to local galaxies.  
Thus, we conclude that it is premature to conclude that $z\sim2$ star--forming galaxies 
have extremely high densities and ionization parameters, 
since the best--measured example, \arcname, does not.

We measure the relative abundances of N, Ne, and Ar compared to O.  
The Ne/O and Ar/O ratios are solar with uncertainties of  $\pm 0.14$ dex, which 
is reassuring since all these elements are alpha--process and should enrich in lockstep.
We believe this to be the first time the Ne/O ratio has been measured at $z\sim2$.  
The N/O ratio is one dex below the solar value, indicating that secondary N production 
has not yet begun in earnest.
The O abundance and N/O ratio are startlingly similar to those of cB58; it is not clear whether this 
agreement is merely coincidental, or indicates characteristic values for star--forming galaxies at this epoch.

\section{Future Directions}  This is by no means the last word on \arcname\ or on diagnostic spectroscopy of
lensed galaxies.  The following observations should significantly increase what can be learned
about the physical conditions of \arcname.
First, a direct measurement [O III]~4363, or an even stricter upper limit, should provide 
a more stringent test of the bright-line O abundance diagnostics.  This comparison, and a host of 
other constraints, are limited by the current uncertainty in the measured extinction.  This would
be improved by a deeper integration of H$\beta$/H$\gamma$ 
or a simultaneous measurement of H$\alpha$/H$\beta$.  

Thus far, we have considered only the spatially--integrated spectrum across the brightest portion of the arc.  
It will be fascinating to spatially map the physical conditions across this portion, and the 
fainter sections of the arc, to see how widely these physical parameters vary across the galaxy.  Of course, such work
requires a better lensing model, which will be enabled by pending HST observations. 

Finally, we humbly remember that a single galaxy can be a maverick, and that 
only by repeating this work in a representative sample of lensed galaxies 
will the physical conditions of star formation at this epoch be confidently characterized.  
Larger samples will also fill in the BPT and 
mass--metallicity relations, to characterize the scatter in these relations and explore the 
reasons for the scatter.

\acknowledgments
Acknowledgments:  
We thank the IRTF Spex team for making public their telluric correction routine and their
tool to find telluric standard stars, at http://irtfweb.ifa.hawaii.edu/\~spex/.
We thank Kevin Schawinski for code to generate the SDSS contours in figure~\ref{fig:BPT}, which is
adapted from \citet{kevin}; 
we thank Fuyan Bian for code to generate Figure~\ref{fig:massZ}, which is adapted from \citet{bian}.
JRR gratefully acknowledges the financial support and intellectual freedom of a Carnegie Fellowship.  

Data presented herein were obtained at the W.M. Keck Observatory from telescope time allocated 
to the National Aeronautics and Space Administration through the agency's scientific partnership 
with the California Institute of Technology and the University of California. The Observatory was 
made possible by the generous financial support of the W.M. Keck Foundation.
We acknowledge the very significant cultural role and reverence that the summit 
of Mauna Kea has always had within the indigenous Hawaiian community.  We are most fortunate 
to have the opportunity to conduct observations from this mountain.




\begin{figure}
\figurenum{3}
\includegraphics[width=3in,height=4in,angle=0]{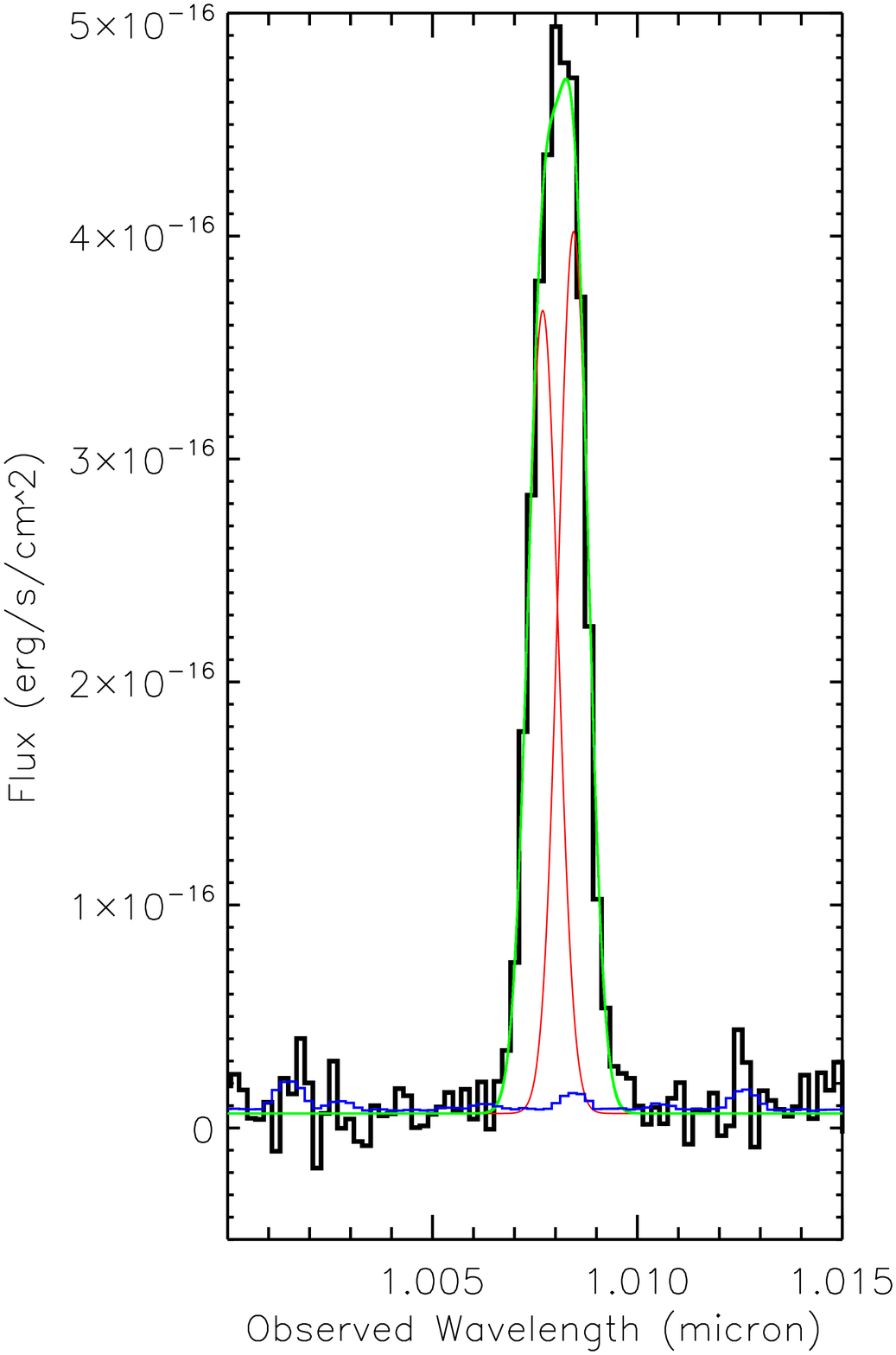}
\figcaption{Two-component Gaussian fit to the [O II] 3726, 3729 doublet.  
Red lines show the fit to each emission line; a green line shows the summed fit.  
Blue shows the $1\sigma$ error spectrum.}
\label{fig:fit3727}
\end{figure}

\begin{figure}
\figurenum{4}
\includegraphics[width=4in,angle=270]{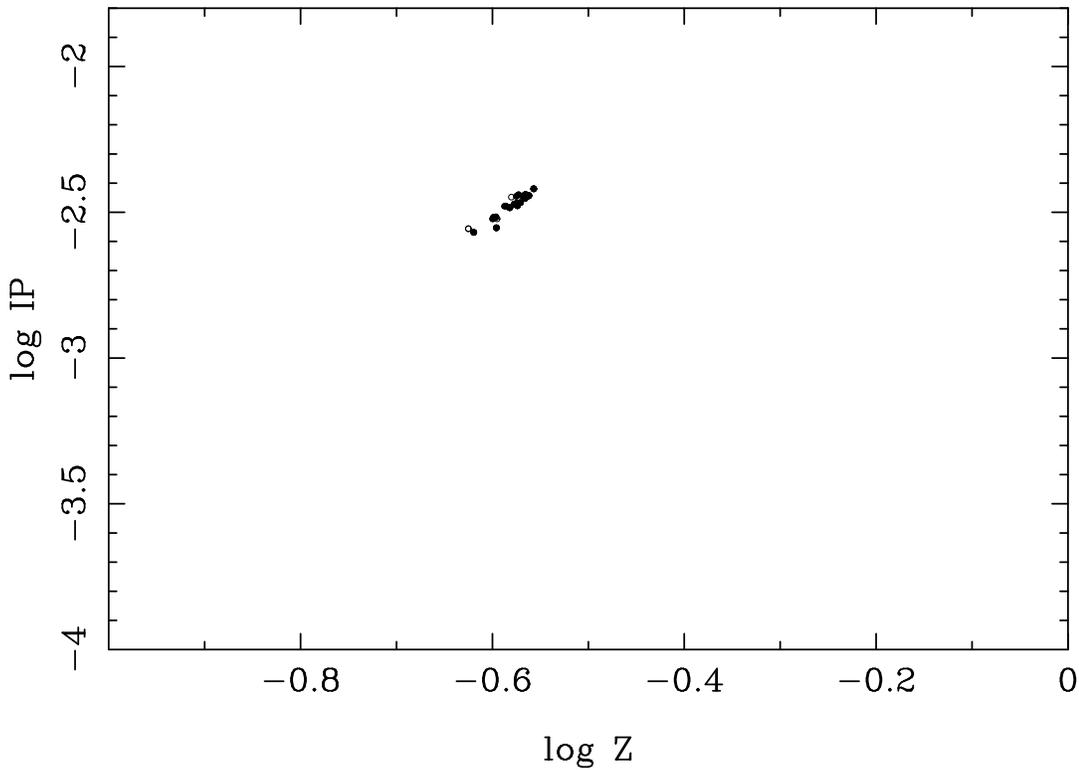}
\figcaption{Starburst99 + Cloudy models, for constant star formation.  
A grid of models was run, varying the electron density through $\pm 1\sigma$ of the value  
measured in \S\ref{sec:ne}, and the starburst age.  The input spectrum changes 
slightly due to stochastic effects.  The filled circles show models with the 
density measured assuming $T_e = 1.2 \times 10^4$~K, 
and open circles show models using the density measured assuming $T_e = 10^4$~K.
The measured spectral lines and the density measurement tightly constrain the 
ionization parameter and metallicity.  
}
\label{fig:cloudy}
\end{figure}

\clearpage 

\begin{figure}
\figurenum{5}
\includegraphics[width=3.5in,angle=270]{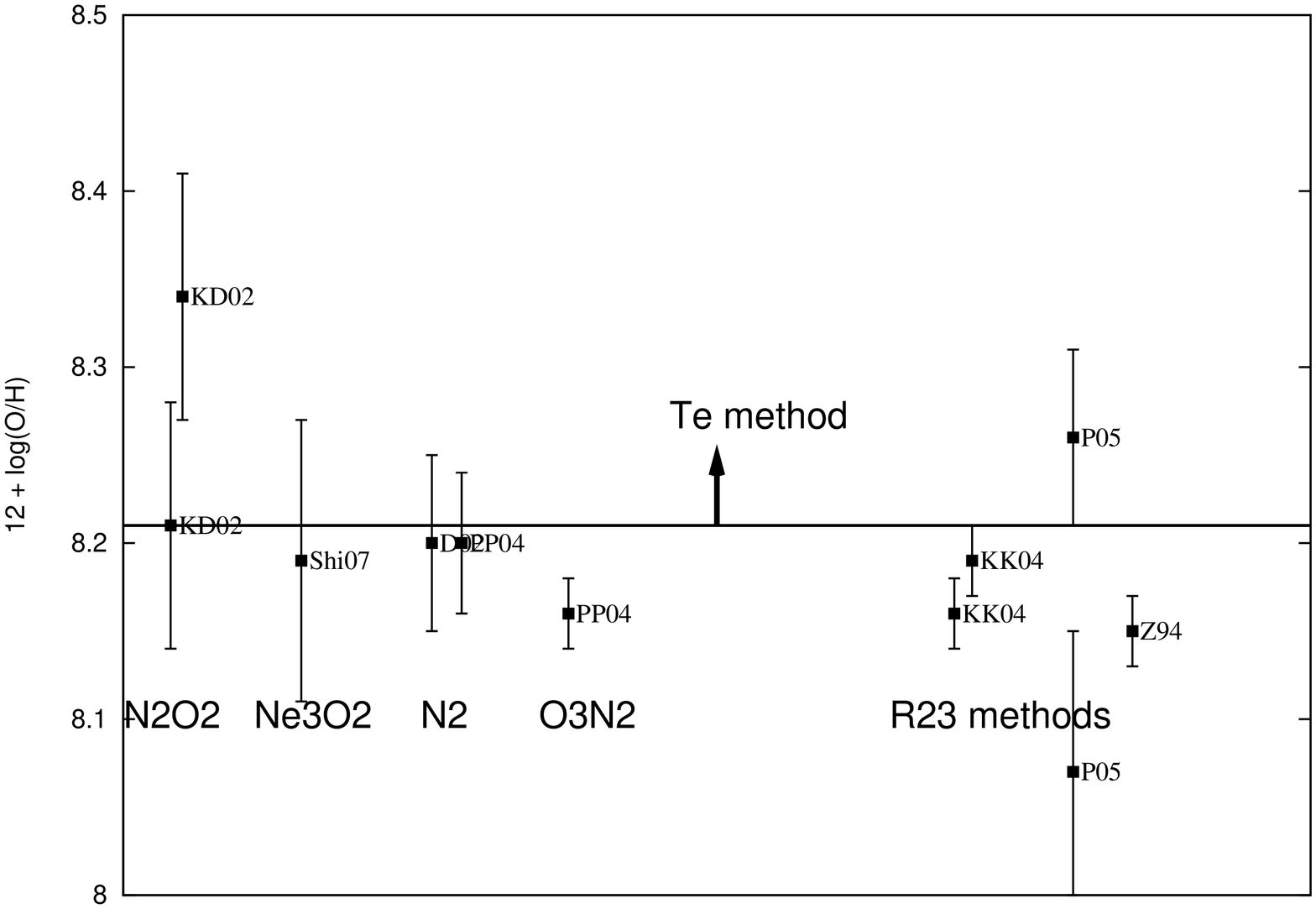}

\includegraphics[width=3.5in,angle=270]{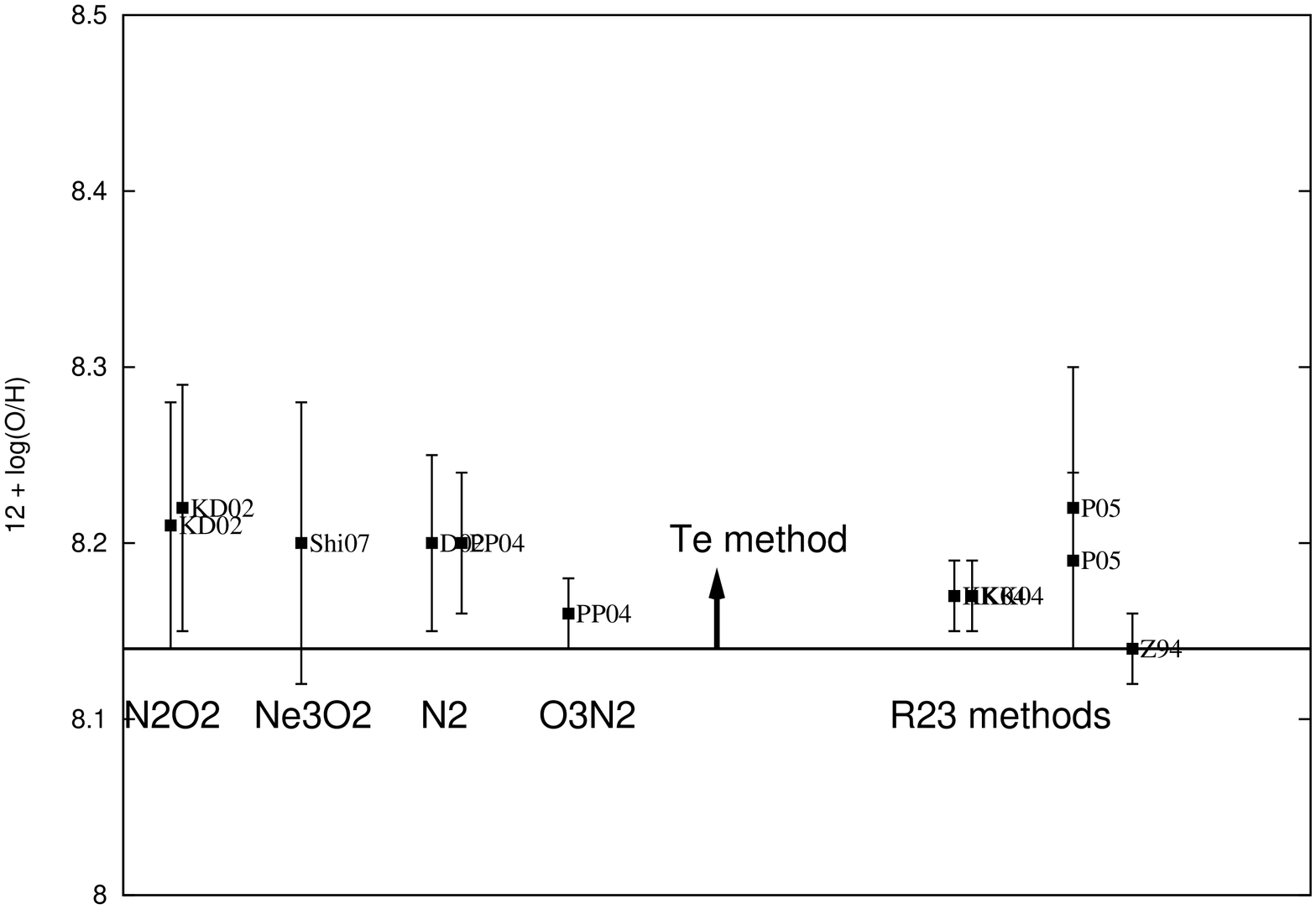}
\figcaption{Comparison of oxygen abundance diagnostics for \arcname. 
The top panel assumes $A_v=0$, and the bottom panel assumes $A_v=0.7$.  
The relative calibration offsets observed at z$=$0 \citep{KE08} have already been removed; all 
indices are on the relative frame of N2 in \citet{pettinipagel}.
}
\label{fig:Zs}
\end{figure}

\begin{figure}
\figurenum{6}
\includegraphics[width=4in,angle=0]{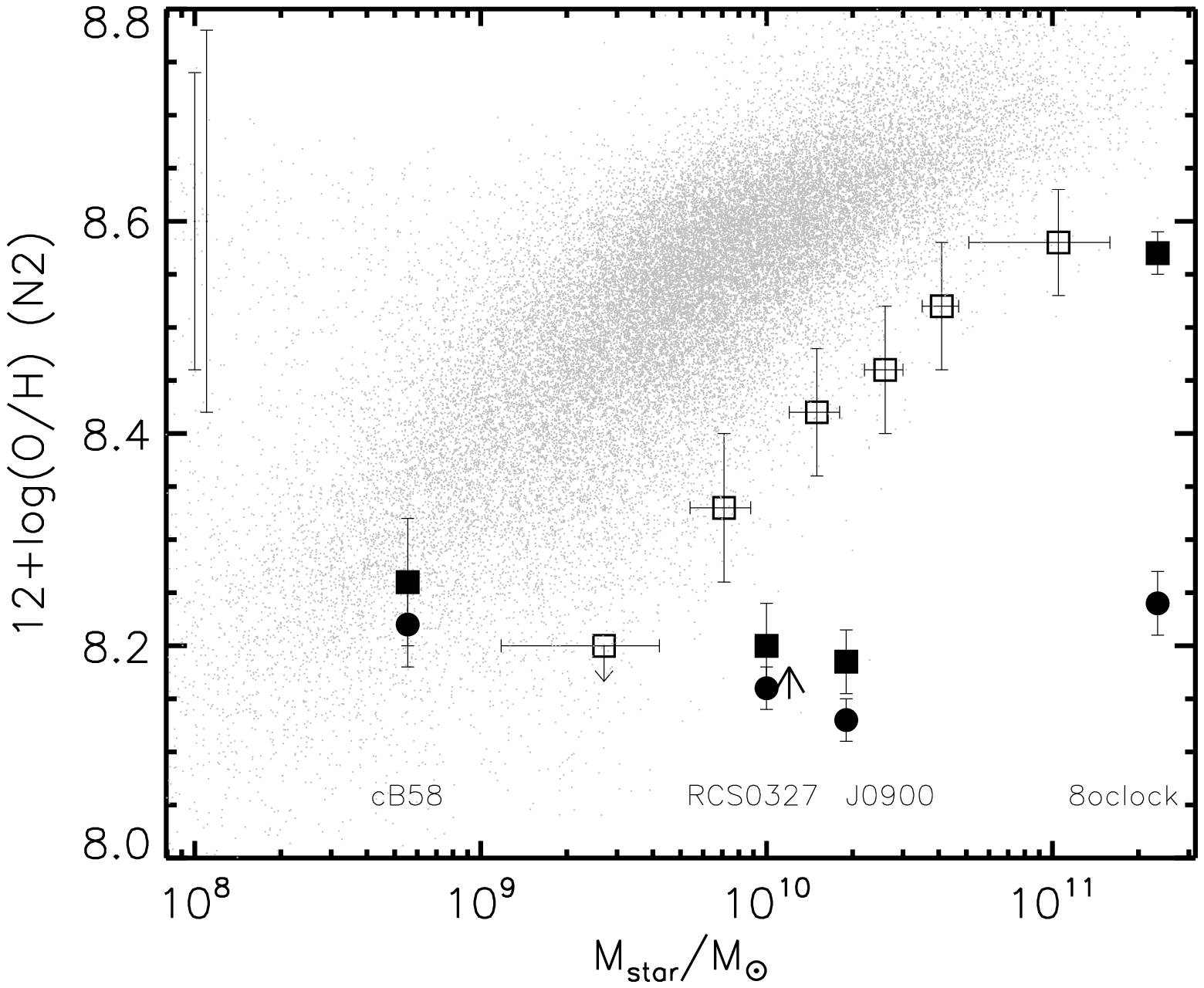}
\figcaption{The Mass--Metallicity relation at z$=$0 and z$=$2.
SDSS galaxies define the z$=$0 relation  \textit{(grey cloud)}.   
The z$=$2 relation is shown by stacked spectra of unlensed z$=$2 LBGs from \citet{erb06} \textit{(hollow squares)}, 
as well as four lensed galaxies at z$\sim$2:
J0900+2234 from \citet{bian}; 
cB58 from \citet{teplitz} and \citet{siana_cb58}; 
the 8 o'clock arc from \citet{finkelstein}; 
and \arcname\ from this work \textit{(filled points)}.   
For consistency, when authors fit stellar masses using a Salpeter IMF, we have converted to a 
\citet{chabrier03} IMF by dividing M$_*$ by 1.8.  
Squares and circles show abundances determined from the N2/H$\alpha$ and O3N2 calibrations of
\citet{pettinipagel}, respectively, where the O3N2 abundances have been brought onto the N2/H$\alpha$
relative system using the small conversion of \citet{KE08}. 
An arrow shows the lower limit on abundance derived from the T$_e$ method for \arcname.
Errorbars on the data show the propagated flux uncertainty.  The errorbars at upper left show the 
$1\sigma$ calibration uncertainties of \citet{pettinipagel}.  
}
\label{fig:massZ}
\end{figure}

\begin{figure}
\figurenum{7}
\includegraphics[width=4in,angle=90]{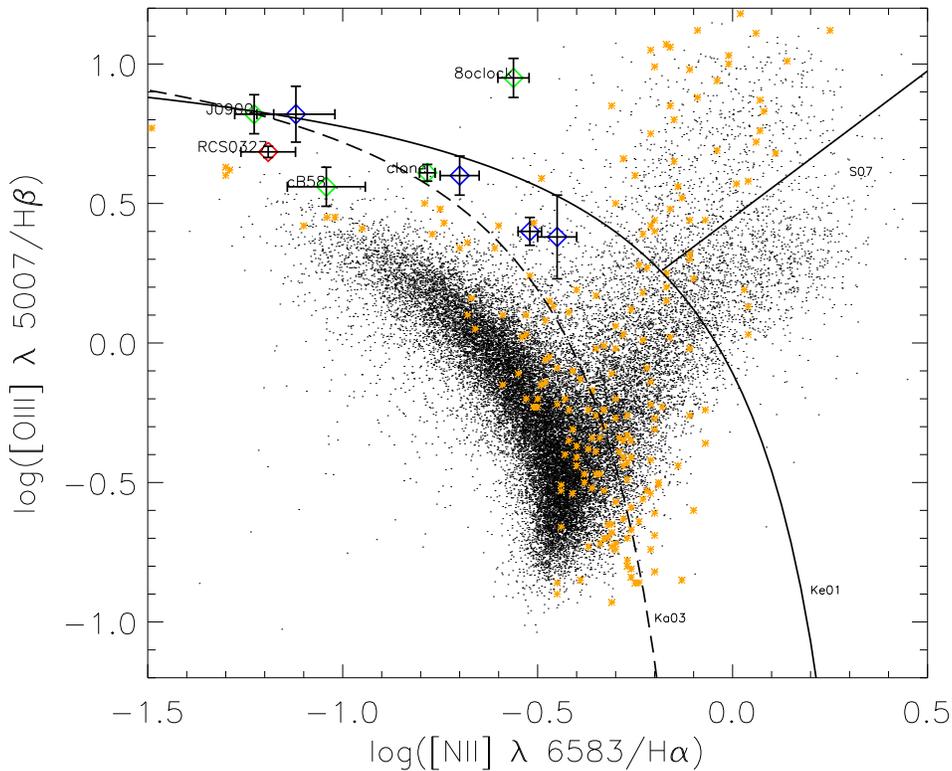}
\figcaption{The BPT diagram.   
The z$=$0 relation is defined by the SDSS DR7 spectroscopic sample 
(\citealt{dr7}; black dots),  
and by the IR--bright galaxies of \citet{Kewley_IR} \textit{(orange asterisks)}.
Stacked samples of galaxies at z$=$2 are from \citet{erb06} \textit{(blue diamonds)}; 
lensed z$=$2 galaxies are from \citet{bian}, \citet{teplitz}, and \citet{finkelstein} 
\textit{(green diamonds)},  and this work\textit{ (red diamond)}.
}
\label{fig:BPT}
\end{figure}

\end{document}